\documentclass[reprint, amsmath,amssymb, aps,floatfix]{revtex4-2}
\usepackage{graphicx}
\usepackage{dcolumn}
\usepackage{bm}
\usepackage{leftidx}
\usepackage[utf8]{inputenc}
\usepackage{txfonts}
\usepackage{diagbox}
\usepackage{slashbox}
\usepackage{xcolor}
\usepackage{enumitem}
\usepackage{floatrow}

\usepackage[breaklinks=true]{hyperref}

\definecolor{dark-green}{rgb}{0,0.7,0}
\definecolor{dark-blue}{rgb}{0,0.1,1}
\definecolor{med-blue}{rgb}{0,0.7,1}
\definecolor{mblue}{rgb}{0,0.2,1}
\definecolor{midnightblue}{rgb}{0.1,0.1,0.4}
\definecolor{skyblue}{rgb}{0.35,0.81,0.9}
\definecolor{steelblue}{rgb}{0.39,0.72,1}

\definecolor{dred}{rgb}{0.7,0,0}
\definecolor{dark-red}{rgb}{0.5,0,0}
\definecolor{light-red}{rgb}{1,0.8,0.8}
\definecolor{dark-yellow}{rgb}{1,0.8,0}
\definecolor{light-blue}{rgb}{0.8,0.85,1}
\definecolor{light-yellow}{rgb}{1,1,0.784}
\definecolor{grey}{gray}{0.8}

\definecolor{llight-red}{rgb}{0.98,0.933,1}
\definecolor{llight-blue}{rgb}{0.796,1,1}

\definecolor{gray1}{gray}{0.1}
\definecolor{gray2}{gray}{0.2}
\definecolor{gray3}{gray}{0.3}
\definecolor{gray4}{gray}{0.4}
\definecolor{gray5}{gray}{0.5}
\definecolor{gray6}{gray}{0.6}
\definecolor{gray7}{gray}{0.7}
\definecolor{gray8}{gray}{0.8}
\definecolor{gray9}{gray}{0.9}

\definecolor{darkblue}{rgb}{0,0,.7}
\hypersetup{colorlinks=true, breaklinks=true, linkcolor=darkblue, menucolor=darkblue, urlcolor=darkblue, citecolor=darkblue}

\begin{document}

\preprint{APS/123-QED}

\title{Vacuum polarization effects in the background of a deformed compact object and implications for photon velocity}

\author{Daniel Amaro}
 \email{daniel.amaro@zarm.uni-bremen.de}
  \affiliation{Center of Applied Space Technology and Microgravity (ZARM), University of Bremen, 28359 Germany} 
  \affiliation{Physics Department, Universidad Autónoma Metropolitana-Iztapalapa, 09340, CDMX, México}
 \author{Shokoufe Faraji}
 \email{s3faraji@uwaterloo.ca}
\affiliation{Waterloo Centre for Astrophysics, University of Waterloo, Waterloo, ON N2L 3G1, Canada} 
    \affiliation{Department of Physics and Astronomy, University of Waterloo, Waterloo, ON N2L 3G1, Canada}
\affiliation{Perimeter Institute for Theoretical Physics, 31 Caroline Street North, Waterloo, N2L 2Y5, Canada}
\affiliation{Center of Applied Space Technology and Microgravity (ZARM), University of Bremen, 28359 Germany
}%

\begin{abstract}

This paper studies the impact of vacuum polarization on light propagation in the background of a distorted, deformed compact object. Focusing on a spacetime containing two quadrupole parameters associated with the central object and external fields, we explore how these parameters influence observable effects as dynamical degrees of freedom. In this setup, we investigate electromagnetic birefringence, noting distinct polarization-dependent photon velocity variations and gravitational lens effects. Although current resolution may limit detection, future high-precision observations could reveal these quantum electrodynamics (QED) induced birefringence effects, advancing our understanding of vacuum birefringence in astrophysical contexts. We further analyze the dependence of shadow properties on the model's variables, using observational data from Sgr A*. 
\end{abstract}

\maketitle


\section{Introduction}

The exploration of quantum field theory in curved spacetime is of paramount significance as it seeks to unify the principles of quantum mechanics with the framework of general relativity. This interdisciplinary attempt has yielded several profound predictions. One of the cornerstone predictions is particle creation, a phenomenon where the curvature of spacetime leads to the spontaneous generation of particles from vacuum fluctuations. This effect is particularly prominent in extreme environments, such as near black holes or during the epoch of cosmic inflation. Among these predictions, Hawking radiation stands out as one of the most famous, which theorizes that black holes can emit thermal radiation due to quantum effects near the event horizon, leading to the eventual evaporation of the black hole itself. Vacuum polarization in curved spacetime suggests that the vacuum state of a quantum field can become polarized due to the influence of gravitational fields. This polarization can lead to significant observable effects, such as alterations in the electromagnetic properties of particles or the emergence of an effective cosmological constant. Another critical concept is backreaction, which posits that quantum fields can influence the geometry of spacetime through their energy-momentum tensor. This interaction can result in modifications to the curvature and dynamics of spacetime, with potential implications for the evolution of cosmological models and the formation and behavior of black holes. As a result, quantum field theory in curved spacetime continues to be an active and exciting area of research, offering profound insights into the fundamental nature of the universe. It is important to note that while these predictions have been extensively supported by theoretical calculations, some aspects still require further experimental or observational evidence.

Aside from these main predictions, quantum field theory in curved spacetimes allows for interesting phenomena and effects like superluminal photon propagation which first was introduced in \cite{PhysRevD.22.343} in the Schwarzschild, de Sitter and Robertson-Walker spacetimes. In \cite{1981PThPh..65.1058O} the massless neutrino propagating in Robertson-Walker spacetime has been studied. Soon after photon propagation following \cite{PhysRevD.22.343} was studied in Reissner-Nordström metric \cite{1994NuPhB.425..634D}, Kerr metric \cite{1996PhLB..367...75D}, and Reissner-Nordström Anti-de Siter background \cite{1998NuPhB.524..639C}. More recently, studies in other spacetimes have been carried out \cite{2014PhRvD..89j4014C, JING2016219}. One approach to studying this phenomenon is through effective actions or effective equations of motion that capture the relevant physics at different energy scales. In the pioneering work \cite{PhysRevD.22.343}, Drummond and Hathrell considered the impact of one-loop vacuum polarization on a general gravitational background in Quantum Electrodynamics (QED) and its contribution to the photon effective action. 

The primary finding of this study is that quantum corrections introduce tidal gravitational forces on photons, which can modify their propagation characteristics. Under specific conditions of motion and polarization, photons may travel at speeds exceeding the speed of light, $c$. However, when vacuum polarization is neglected, we can deduce the characteristics of photon propagation based on the equivalence principle. Moreover, observations from 2018 of visible polarization from a radio-quiet neutron star provide strong evidence that vacuum birefringence is influencing the polarization of photons \cite{10.1093/mnras/stw2798}. These observations underscore the importance of considering quantum corrections in our understanding of photon propagation in curved spacetime, further emphasizing the intricate interplay between quantum mechanics and general relativity.


Further analysis in this direction also considers the effects on gravitational lensing \cite{2015JCAP...10..002C, 2016EPJC...76..357L, 2018EPJC...78..191C, 2020PhRvD.101l4038B, 2018AnPhy.399..193O, 2020ChPhC..44i5105A, 2021EPJC...81..991Z, 2023Univ....9..130A, chen2023kerr}. Gravitational lensing studies often employ the thin-lens approach, which assumes the lens is a perturbation of the background spacetime, with light bending occurring primarily in the lens plane, while light rays travel as straight lines elsewhere. In the weak-field thin-lens approximation, only small deflection angles are considered. However, in strong gravitational fields, such as those produced by black holes, light rays can bend significantly and may even orbit the lens multiple times. The strong-field thin-lens model \cite{2000PhRvD..62h4003V, 2002PhRvD..66j3001B}, describing the bending angle approaching infinity as rays approach the unstable circular orbit, is the most widely used method for strong gravitational lensing. This model accounts for the extreme deflection of light in the vicinity of compact massive objects, providing a comprehensive framework for understanding the complex interactions between light and gravity in extreme environments. 

However, both weak and strong-field thin-lens models assume that the source and observer are at infinity, which is often not the case in real astrophysical settings. The most general approach, developed by Frittelli et al. \cite{PhysRevD.59.124001, PhysRevD.61.064021}, is based on an exact study of the null geodesics followed by light rays. In the exact-lens approach, bending does not occur solely in the lens plane, and the positions of the observer and source are not restricted to infinity. This method allows for a comprehensive examination of the effects of strong gravitational fields on light from sources very close to the lens. Moreover, this approach facilitates the analysis of modifications to light propagation due to strong gravitational influences. 
By considering the exact trajectories of light rays in curved spacetime, this framework provides a more accurate and versatile understanding of gravitational lensing, accommodating a wider range of astrophysical conditions and enabling the study of more complex lensing scenarios.


In this paper, we investigate the impact of vacuum polarization on photon propagation and extend this approach to the background of a distorted, deformed compact object characterized by quadrupoles. Description of this spacetime is provided in section \ref{metricsec}. Our study focuses on the analogous problem mentioned above and employs the one-loop approximation. We found that the propagation of photons can be influenced by polarization, and under certain conditions, their speed of propagation can exceed the speed of light. Additionally, we analyzed the motion of polarized photons and discussed the effects of nonminimal coupling on gravitational lensing.

The outline of the paper is as follows: In Section \ref{polar}, we present the photon propagation equation. Section \ref{metricsec} briefly introduces the background spacetime. The propagation equation is derived in Section \ref{bif}. Gravitational birefringence and polarizations are discussed in Section \ref{polarsec}. In Section \ref{lens}, we examine the birefringence effect on lensing. Finally, the summary and conclusions are presented in Section \ref{sum}. Throughout this work, Greek indices $\mu = t, x, y, \phi$ are used for the coordinate basis, and Latin indices $a = 0, 1, 2, 3$ denote components in the orthonormal frame. The dot denotes derivatives with respect to the affine parameter, and prime denotes derivatives with respect to the radial coordinate, unless otherwise stated.

\section{Photon propagation} \label{polar}

The action of the electromagnetic field in curved spacetime for low frequencies can be written as
\begin{equation}
W_{\text{EM}} =-\frac14 \int dx^4\sqrt{-g}F_{\mu\nu}F^{\mu\nu}
+W_1\, ,
\end{equation}
where $F_{\mu\nu}=\partial_{\mu}A_{\nu}-\partial_{\nu}A_{\mu}$ is the Faraday tensor and $g$ is the determinant of the metric $g_{\mu\nu}$. The presence of $\sqrt{-g}$ on
the curved background, is due to the fact that the field equations must be invariant under arbitrary smooth transformations of the coordinates. The second term is the nonminimally coupled sector $W_1$ in the $RF^2$ form that explicitly breaks the conformal invariance of $U(1)$ through gravitational couplings 
\begin{eqnarray}
W_1&=& \frac{1}{m^2}
\int dx^4\sqrt{-g}\left(aRF_{\mu\nu}F^{\mu\nu}
+bR_{\mu\nu}F^{\mu\sigma}F^{\nu}_{\ \sigma}\right.\nonumber\\
&&\qquad \left. cR_{\mu\nu\sigma\tau}F^{\mu\nu}F^{\sigma\tau}
+dD_{\mu}F^{\mu\nu}D_{\sigma}F^{\sigma}_{\ \nu}\right)\, ,
\end{eqnarray}
where $A_{\mu}$ is the electromagnetic four potential. However, Gauge invariance implies that $W_1$ depends on $F_{\mu\nu}$ rather than directly on $A_{\mu}$. The operator $D_{\mu}$ denotes the covariant derivative in curved spacetime, defined in terms of the connection coefficient $\Gamma^\lambda_{\mu \nu}$. Also, $R_{\mu\nu\sigma\rho}$ is the Riemann curvature tensor, and $R_{\mu\nu}=R^{\sigma}_{\mu\sigma\nu}$ and $R=g^{\mu\nu}R_{\mu\nu}$ are respectively the Ricci tensor and the scalar curvature. The coefficients $a$, $b$, $c$, and $d$ are coupling constants. In this context, the parameter $m$ represents a squared mass scale that ensures dimensional consistency in most interacting quantum field theories, excluding conformal field theories. This term arises due to one-loop vacuum polarization effects in curved spacetimes and can be as small as the electron mass $m_e$, since vacuum polarization refers to the temporary existence of a photon as a virtual pair of an electron and a positron. This transition grants the photon a size proportional to the Compton wavelength of the electron, denoted as $\lambda_c = 1/m_e$. Consequently, the curvature of the gravitational field can influence the motion of the photon \footnote{However, this is not in favor of the primeval fields scenario and therefore inflation which is a prime candidate for the production of primeval magnetic fields. This scenario requires $r > 10^{-10}$ for the scales of astrophysical interest, $\lambda$ around Mpc. However, considering the mentioned approach the primeval fields produced are typically small $r \sim 10^{-68}$.}. In addition, due to the interacting terms between the electromagnetic field and the spacetime curvature, the principle of equivalence is lost and superluminal photons are allowed in curved spacetime, and therefore do not violate the causality. This phenomenon arises because the dispersion relation for photons in curved spacetime can be modified by the interaction terms between the electromagnetic field and the curvature tensor $R_{\mu \nu \sigma \rho. }$ These modifications can result in an effective refractive index that allows phase velocities exceeding the speed of light in vacuum. However, the principle of causality is preserved because the group velocity, which governs the transport of information and energy, remains within the causal light cone of the underlying spacetime metric. The apparent superluminal behavior reflects a shift in the local propagation characteristics of the electromagnetic wave, influenced by the spacetime geometry and higher-order corrections. Importantly, the causal structure of the spacetime remains intact, as no physical signal or information is transmitted faster than the universal speed limit dictated by general relativity.

The electromagnetic field equation derived from this action $W_{\text{EM}}$ can then be written as

\begin{eqnarray}
&&D_{\mu}F^{\mu\nu}+\frac{1}{m_e^2}
D_{\mu}\left[4aRF^{\mu\nu}
+2b\left(R^{\mu}_{\ \sigma}F^{\sigma\nu}-R^{\nu}_{\ \sigma}F^{\sigma\mu}\right)\right.\nonumber\\
&&\left.\hspace{2.5cm}+4cR^{\mu\nu}_{\ \ \sigma\tau} F^{\sigma\tau}\right]=0\, .\label{modif}
\end{eqnarray}
The values for the coupling constants $a, b$ and $c$ can be calculated by considering the coupling of a graviton to two on-mass-shell photons in the flat-space limit \cite{PhysRevD.22.343} as

\begin{align}
    \{a,b,c\}=-\frac{\alpha}{720\pi}\{5,-26,2\}\, .
\end{align}
The term with coefficient $d$
could be omitted since it only influenced the motion as a second-order correction. Furthermore, for fields that satisfy the Einstein vacuum equations, $R_{\mu\nu}=0$, the equation of motion is just determined by the coupling with the Riemann curvature tensor,
\begin{eqnarray}
D_{\mu}F^{\mu\nu}+\xi^2R^{\mu\nu}_{\ \ \sigma\tau}
D_{\mu}F^{\sigma\tau}&=&0\, ,\label{vacuum0}
\end{eqnarray}
since the photon is treated as a test particle, its effect on the spacetime is negligible. If one aims to study QED corrections to the photon propagation, the value of $\xi^2=\tilde{\alpha}_{\text{f}}/(90\pi m_e^2)$ is fixed, with $\tilde{\alpha}_{\text{f}}$ the fine structure constant, otherwise one can consider it as a free parameter. Furthermore, in the case of vacuum field equation it has been considered as the coupling constant with which photons couple to the Weyl tensor (see e.g., \cite{2014PhRvD..89j4014C}). In the following, we consider $\xi^2\ll1$ as an infinitesimal parameter and we omit higher order terms ${\cal{O}}(\xi^4)$; the terms involving $\xi^2$ correspond to corrections due to the nonminimal coupling of gravity and electrodynamics.

Additionally to (\ref{vacuum0}), the electromagnetic equation is given by
\begin{equation}\label{BianchiMax}
D_{\rho}F_{\mu\nu}+D_{\mu}F_{\nu\rho}+D_{\nu}F_{\rho\mu}=0\, .
\end{equation}
To derive the equation governing the characteristics of photon propagation, we employ the geometrical optics plane wave approximation, utilizing both the electromagnetic equations \eqref{vacuum0} and \eqref{BianchiMax} and set $F_{\mu\nu}=f_{\mu\nu}e^{i\theta}$, where $f_{\mu\nu}$ is a slowly varying amplitude with respect to a rapidly varying phase $\theta$. If we consider $k_{\mu}=D_{\mu}\theta$ and ignore higher-order derivatives, then equations \eqref{vacuum0} and \eqref{BianchiMax} are respectively rewritten as
\begin{eqnarray}
k_{\mu}f^{\mu\nu}+\xi^2 R^{\mu\nu}_{\ \ \sigma\tau}k_{\mu}f^{\sigma\tau}&=&0
\, ,\label{vacuum}\\
k_{\rho}f_{\mu\nu}+k_{\mu}f_{\nu\rho}+k_{\nu}f_{\rho\mu}&=&0\, . \label{Bianchi}
\end{eqnarray}
From equation \eqref{Bianchi} one can deduce 
\begin{align} \label{fmain}
    f_{\mu\nu}=k_{\mu}a_{\nu}-k_{\nu}a_{\mu},
\end{align}
where $a_{\mu}$ is the polarization vector satisfying the condition $k_\mu a^\mu=0$. Considering that the amplitude $f_{\mu\nu}$ has three independent components, contracting equation (\ref{Bianchi}) with $k^{\nu}$ gives, 

\begin{align}
    k^2f^{\mu\nu}=k_{\alpha}f^{\alpha\nu}k^{\mu}-k_{\alpha}f^{\alpha\mu}k^{\nu}\, .
\end{align}
Finally, by combining this with equation \eqref{vacuum} and using the Bianchi identity, the propagation equation reads
\begin{eqnarray} \label{main2}
k^2f^{\mu\nu}+\xi^2k_{\alpha}\left(k^{\mu}R^{\alpha\nu}_{\ \ \sigma\tau}
-k^{\nu}R^{\alpha\mu}_{\ \ \sigma\tau}\right)f^{\sigma\tau}=0\, .\label{vacuum2}
\end{eqnarray}
In the following, we discuss the implications of these equations for photon propagation in the generalized q-metric background. 

\vspace{0.5 cm}
\section{generalized q-metric} \label{metricsec}
In general relativity, Weyl's class of solutions constitutes a set of exact solutions to the vacuum Einstein field equations that are characterized by their static and axisymmetric properties \cite{doi:10.1002/andp.19173591804}. Among these solutions, the $\rm q$-metric stands out as the simplest and most analytically applicable generalization of the Schwarzschild family. It describes static, axially symmetric, and asymptotically flat solutions that include quadrupole moments to account for deviations from spherical symmetry. The $\rm q$-metric represents the gravitational field outside an isolated compact object and has been extensively studied using various approaches. Additionally, this metric has further been broadened by considering the distribution of mass in its vicinity, thus relaxing the assumption of isolation \cite{universe8030195}. This generalized $\rm q$-metric can be understood as adding additional external gravitational fields, analogous to introducing a magnetic environment \cite{1976JMP....17...54E}. Notably, the presence of a quadrupole moment can significantly alter the geometric properties of spacetime. The metric expressed in prolate spheroidal coordinates is given by \cite{universe8030195}
\begin{eqnarray}\label{metric}
ds^2&=&-\left(\frac{x-1}{x+1}\right)^{(1+\alpha)}e^{2\psi}dt^2
+M^2(x^2-1)e^{-2\psi}\left(\frac{x+1}{x-1}\right)^{(1+\alpha)}\nonumber\\
&&\hspace{-1cm}\times\left[\left(\frac{x^2-1}{x^2-y^2}\right)^{\alpha(2+\alpha)}e^{2\chi}
\left(\frac{dx^2}{x^2-1}+\frac{dy^2}{1-y^2}\right)+(1-y^2)d\phi^2\right]\, ,\nonumber\\
\end{eqnarray}
where the functions $\psi=\psi(x,y)$ and $\chi=\chi(x,y)$ present the influence of the external matter encoded by a set of multipole moments. However, up to the dominant term, the quadrupole moments $\beta$, they read as
\begin{eqnarray}
\psi(x,y)&=&-\frac{\beta}{2}\left(-3x^2y^2+x^2+y^2-1\right)\, ,\label{psi}\\
\chi(x,y)&=&-2x\beta\left(1+\alpha\right)\left(1-y^2\right)\, \nonumber\\
&&\hspace{-1cm} +\frac{\beta^2}{4}\left(x^2-1\right)\left(1-y^2\right)
\left(-9x^2y^2+x^2+y^2-1\right)\, .\label{chi}
\end{eqnarray}
The deformation parameter $\alpha$ is connected to the compact object’s deformation, while the distortion parameter $\beta$ is related to an additional external gravitational field, like an external mass distribution or a magnetic surrounding. While both parameters are relatively small, the maximum and the minimum allowed values of $\beta$ are dependent on the chosen value of $\alpha$ \cite{universe8030195}. The relation between this coordinate system $(t, x, y, \phi)$ and the $(t, r, \theta, \phi)$ coordinates is given by

\begin{align}\label{transf1}
 x =\frac{r}{M}-1 \,, \quad  y= \cos\theta\,.
\end{align}
To gain an intuitive understanding of the role of $\beta$, we can consider Newtonian gravity, and denote the quadrupole moment by $\beta_N$. In Newtonian theory, it is well-known that a multipole expansion dominated by a quadrupole moment $\beta_N$ can be modeled by two equal point-like masses, $m$, located along an axis, such as the $z$-axis, at some distance from the center. Additionally, consider an infinitesimally thin ring with mass $M$ and radius $R$ situated in the plane perpendicular to this axis. If the gravitational field contribution from the point-like masses is greater than that from the ring, then $\beta_N<0$. Conversely, if the ring's contribution is greater, then $\beta_N>0$. When $\beta_N<0$, a net force is directed toward the $z$-axis, creating a potential barrier. On the other hand, if $\beta_N>0$, a net force is directed toward the ring, outward from the central object, balancing the gravitational pull of the central source and the external fields.

\vspace{0.5cm}

\section{Photon propagation in generalized q-metric} \label{bif}
In this section, we study the motion of the photons in the equatorial plane, $y=0$ (equivalently $\theta=\frac{\pi}{2}$), of the background spacetime. We start with the Minkowski metric $ds^2=\eta_{ab}\omega^{a}\omega^{b}$, with $\eta_{ab}=\text{diag}\{-1,1,1,1\}$, and the orthonormal tetrad $\omega^{a}=e_{\mu}^{a}\textbf{d}x^{\mu}$, where
\begin{eqnarray}
e_{\mu}^{a}=\text{diag}\left\{\sqrt{-g_{tt}},\sqrt{g_{xx}},
\sqrt{g_{yy}},\sqrt{g_{\phi\phi}}\right\}\, .\label{tetrad}
\end{eqnarray}
Additionally, since the values of $\beta$ are also very small when multiplied by $\xi^2$ we can neglect the second-order terms in $\xi^2 \beta$. The components of the Riemann tensor in the orthonormal basis read [Appendix \ref{AppendixA}]
\begin{eqnarray}
R^{01}_{\ \ 01}=R^{23}_{\ \ 23}&=& \zeta(x)\left\{(1+\alpha) \left[\alpha(2+\alpha) 
-2(1+\alpha) x + 2 x^2\right]
\right.\nonumber\\ 
&&\hspace{-0.3cm}\left. +\beta x (x^2-1)\left[-3(1+\alpha)^2 +4(1+\alpha)x + x^2\right]\right\} \, ,\nonumber\\ 
R^{02}_{\ \ 02}=R^{13}_{\ \ 13}&=& \zeta(x)\left\{-(1+\alpha) \left[\alpha(2+\alpha)
-(1+\alpha) x + x^2\right]
\right.\nonumber\\ 
&&\hspace{-0.3cm}\left. +\beta x (x^2-1)\left[3(1+\alpha)^2 -2(1+\alpha)x -2x^2\right]\right\} \, ,\nonumber\\ 
R^{03}_{\ \ 03}=R^{12}_{\ \ 12}&=& x\zeta(x)\left[(1+\alpha) (1+\alpha-x)\right.\nonumber\\
&&\left.\qquad+ \beta x (x^2-1)(-2-2\alpha+x)\right]\, ,\nonumber\\ 
R^{01}_{\ \ 02}=R^{02}_{\ \ 01}&=& R^{13}_{\ \ 23}=R^{23}_{\ \ 13}=0\, ,\label{Rabcd}
\end{eqnarray}
with 
\begin{equation} \label{zetax}
\zeta(x)\approx\frac{1+\beta\left[1+4x(1+\alpha)-x^2\right]}{M^2x(x^2-1)^2}
\left(\frac{x^2-1}{x^2}\right)^{-\alpha(2+\alpha)}\left(\frac{x-1}{x+1}\right)^{1+\alpha}\, .
\end{equation}
Hence, the Riemann tensor can be rewritten as 
\begin{eqnarray}
R^{\mu\nu}_{\ \ \sigma\tau}&=&A\left[\delta^{\mu}_{\sigma}\delta^{\nu}_{\tau}
-\delta^{\mu}_{\tau}\delta^{\nu}_{\sigma}\right]
+B\left[U^{\mu\nu}_{01}U_{\sigma\tau}^{01}
+U^{\mu\nu}_{23}U_{\sigma\tau}^{23}\right]\, \nonumber\\
&&\qquad+C\left[U^{\mu\nu}_{02}U_{\sigma\tau}^{02}
+U^{\mu\nu}_{13}U_{\sigma\tau}^{13}\right]\, , \label{Rmnst}
\end{eqnarray}
with the antisymmetric combination of tetrads
\begin{equation}\label{Umunu}
U_{\mu\nu}^{ab}\equiv e_{\mu}^{a}e_{\nu}^{b}-e_{\mu}^{b}e_{\nu}^{a}\, ,
\end{equation}
and functions $A$, $B$, and $C$ given by
\begin{widetext}
\begin{eqnarray}
A(x)&=&\frac{(1+\alpha) (1+\alpha-x)+\beta\left[(1+\alpha)(1+\alpha-x)
\left(1+4x(1+\alpha)-x^2\right)-x(x^2-1)(2+2\alpha-x)\right]}{M^2(x^2-1)^2}\nonumber\\
&& \qquad\times\left(\frac{x^2-1}{x^2}\right)^{-\alpha(2+\alpha)}
\left(\frac{x-1}{x+1}\right)^{1+\alpha}\, ,\nonumber\\
B(x)&=& \left\{\alpha(2+\alpha)-3x(1+\alpha-x)+\beta\left[
\left(\alpha(2+\alpha)-3x(1+\alpha-x)\right)\left(1+4x(1+\alpha)-x^2\right)\right.\right.\nonumber\\
&&\left.\left.-3x(x^2-1)(1+\alpha-2 x)\right]\right\} \frac{(1+\alpha)}{M^2x(x^2-1)^2}
\left(\frac{x^2-1}{x^2}\right)^{-\alpha(2+\alpha)}\left(\frac{x-1}{x+1}\right)^{1+\alpha}\, ,\nonumber\\
C(x)&=&\frac{-\alpha(1+\alpha)(2+\alpha)+\beta\left[3x(x^2-1)(1+\alpha-x)(1+\alpha+x)
-\alpha(1+\alpha)(2+\alpha)\left(1+4x(1+\alpha)-x^2\right)\right]}{M^2x(x^2-1)^2}\nonumber\\
&&\qquad\times\left(\frac{x^2-1}{x^2}\right)^{-\alpha(2+\alpha)}\left(\frac{x-1}{x+1}\right)^{1+\alpha}\, .
\nonumber\\ \label{ABCx}
\end{eqnarray}
\end{widetext}
The tensors $U_{\mu\nu}^{ab}$ satisfy 
\begin{align}
    g_{\alpha\mu}g_{\beta\nu}U^{\alpha\beta}_{0i}&=-U_{\mu\nu}^{0i},\\
    g_{\alpha\mu}g_{\beta\nu}U^{\alpha\beta}_{jk}&=U_{\mu\nu}^{jk}\,.
\end{align}
In the specific case where only deformation is present ($\alpha \neq 0$, $\beta = 0$), the functions described by (\ref{ABCx}) simplify to
\begin{equation}
A(x)
=\frac{(1+\alpha) (1+\alpha-x)}{M^2(x^2-1)^2}
\left(\frac{x^2-1}{x^2}\right)^{-\alpha(2+\alpha)}\left(\frac{x-1}{x+1}\right)^{1+\alpha}\, ,\nonumber
\end{equation}
\begin{eqnarray}
B(x)&=&\frac{(1+\alpha)\left[\alpha(2+\alpha)-3x(1+\alpha-x)\right]}{M^2x(x^2-1)^2}\, \times\nonumber\\
&&\qquad\times \left(\frac{x^2-1}{x^2}\right)^{-\alpha(2+\alpha)}\left(\frac{x-1}{x+1}\right)^{1+\alpha}\, ,\nonumber\\
C(x)&=&-\frac{\alpha(1+\alpha)(2+\alpha)}{M^2x(x^2-1)^2}
\left(\frac{x^2-1}{x^2}\right)^{-\alpha(2+\alpha)}\left(\frac{x-1}{x+1}\right)^{1+\alpha}\, , \quad\label{ABCxb0}
\end{eqnarray}

while in the case where only distortion is present ($\alpha=0$, $\beta\neq0$), they read
\begin{eqnarray}
A(x)&=&-\frac{\left[1+\beta\left(1+6x-x^3\right)\right]}{M^2(x+1)^3}\, ,\nonumber\\
B(x)&=&\frac{3\left[1+\beta x\left(x+5\right)\right]}{M^2(x+1)^3}\, ,\nonumber\\
C(x)&=&-\frac{3\beta(x-1)}{M^2(x+1)}\, .\label{ABCxq0}
\end{eqnarray}

It is straightforward to verify that for the Schwarzschild case, $\alpha=\beta=0$, and in terms of the Schwarzschild-like r-coordinate \eqref{transf1}, one obtains $A=-M/r^3$, $B=3M/r^3$, and $C=0$.


\section{Gravitational birefringence} \label{polarsec}
The concept of birefringence effectively illustrates the difference between how electromagnetic waves behave in media compared to in a vacuum. It shows that the propagation of light is governed by two distinct light cones, which may not necessarily align with the spacetime background. In fact, the field strength components of this system are
physically measurable quantities. Therefore, they have to be determined uniquely. We start by writing the Lorentz components of $f^{\mu\nu}$ which are obtained via contracting $f^{\mu\nu}$ with $U_{\mu\nu}^{ab}$ as

\begin{equation}
f^{ab}=\frac12 f^{\mu\nu}U_{\mu\nu}^{ab}\, .
\end{equation}
The choice of a set of three components that are linearly independent depends on the specific background being studied. For example, contracting \eqref{main2} with the following $U_{\mu \nu}$ components provides a set of three linearly independent equations by defining the vectors
\begin{eqnarray}
l_{\nu}&=&k^{\mu}U_{\mu\nu}^{01}\, ,\nonumber\\
n_{\nu}&=&k^{\mu}U_{\mu\nu}^{02}\, ,\nonumber\\
p_{\nu}&=&k^{\mu}U_{\mu\nu}^{13}\, ,\nonumber\\
m_{\nu}&=&k^{\mu}U_{\mu\nu}^{23}\, . \label{vectors}
\end{eqnarray}
If one considers the above relations and substitutes \eqref{Rmnst} in the propagation equation \eqref{vacuum2}, the electromagnetic field equation is given by
\begin{align}
&\left(1+2\xi^2A\right)k^2f^{\mu\nu}
+2\xi^2B\left(\left[k^{\mu}l^{\nu}-k^{\nu}l^{\mu}\right]f^{01}
+\left[k^{\mu}m^{\nu}-k^{\nu}m^{\mu}\right]f^{23}\right)\nonumber\\
&+2\xi^2C\left(\left[k^{\mu}n^{\nu}-k^{\nu}n^{\mu}\right]f^{02}
+\left[k^{\mu}p^{\nu}-k^{\nu}p^{\mu}\right]f^{13}\right) = 0.
\end{align}
Dividing by $\left(1+2\xi^2A\right)$, and only keeping the first-order correction in $\xi^2$ one obtains
\begin{align}
&k^2f^{\mu\nu}
+2\xi^2B\left(\left[k^{\mu}l^{\nu}-k^{\nu}l^{\mu}\right]f^{01}
+\left[k^{\mu}m^{\nu}-k^{\nu}m^{\mu}\right]f^{23}\right)&&\, \nonumber\\
&+2\xi^2C\left(\left[k^{\mu}n^{\nu}-k^{\nu}n^{\mu}\right]f^{02}
+\left[k^{\mu}p^{\nu}-k^{\nu}p^{\mu}\right]f^{13}\right)=0 \, . \label{emqm}
\end{align}
Now by contracting this equation by each of the tensor components (\ref{Umunu}) we can present this system of equations in a matrix form

\begin{small}
\begin{equation}\label{matrix1}
\begin{pmatrix}
k^2+2\xi^2Bl^2 & 2\xi^2C(l\cdot n) & 2\xi^2C (l\cdot p) & 2\xi^2B(l\cdot m) \\
2\xi^2B(n\cdot l) & k^2+2\xi^2Cn^2 & 2\xi^2C (n\cdot p) & 2\xi^2B(n\cdot m)\\
2\xi^2B(p\cdot l) & 2\xi^2C(p\cdot n) & k^2+2\xi^2Cp^2 & 2\xi^2B (p\cdot m)\\
2\xi^2B(m\cdot l) & 2\xi^2C(m\cdot n) & 2\xi^2C(m\cdot p) & k^2+2\xi^2B m^2
\end{pmatrix}
\begin{pmatrix}
f^{01} \\ f^{02} \\ f^{13} \\ f^{23}
\end{pmatrix}
=0\, .
\end{equation}
\end{small}
This means there is a linear relation between the rows of the matrix. The eigenvalues of this matrix give the light-cone conditions on the photon momentum
\begin{align}
    (\eta_{ab}+\alpha\sigma_{ab})k^ak^b=0\, ,
\end{align}
here $\alpha\sigma_{ab}$ is the one-loop vacuum polarization correction depending on the Riemann curvature tensor at any given point, and the eigenvectors determine the polarizations. The products read as
\begin{align}
&l^2=-k^{(0)}k^{(0)}+k^{(1)}k^{(1)}\nonumber,\\
&n^2=-k^{(0)}k^{(0)}+k^{(2)}k^{(2)}\nonumber,\\ &p^2=k^{(1)}k^{(1)}+k^{(3)}k^{(3)}\nonumber,\\ &m^2=k^{(2)}k^{(2)}+k^{(3)}k^{(3)}\nonumber,\\
&l\cdot n= p\cdot m = k^{(1)}k^{(2)}\nonumber,\\
&l\cdot p= n\cdot m= k^{(0)}k^{(3)}\nonumber,\\
&l\cdot m=n\cdot p=0.\
\end{align}
The $k^{(a)}$ denote the components of the propagation vectors in the orthonormal basis.
Since we are restricting our study to the equatorial plane, $k^{(2)}=0$, one obtains $n\cdot l=l\cdot n=p\cdot m=m\cdot p=0$. Additionally, one can verify that the relations $p\cdot l=-l\cdot p$ and $m\cdot n=-n\cdot m$ are satisfied. Then \eqref{matrix1} becomes
\begin{small}
\begin{equation}\label{matrix2}
\begin{pmatrix}
k^2+2\xi^2Bl^2 & 0 & 2\xi^2C (l\cdot p) & 0 \\
0 & k^2+2\xi^2Cn^2 & 0 & 2\xi^2B(n\cdot m)\\
-2\xi^2B(l\cdot p) & 0 & k^2+2\xi^2Cp^2 & 0\\
0 & -2\xi^2C(n\cdot m) & 0 & k^2+2\xi^2B m^2
\end{pmatrix}
\begin{pmatrix}
f^{01} \\ f^{02} \\ f^{13} \\ f^{23}
\end{pmatrix}
=0\, .
\end{equation}
\end{small}
The determinant of this matrix leads to the light cone condition {\footnote{An alternative way of deriving the dispersion relations is to start from the wave equation in the radiation gauge, where there are components only in the transverse plane, with zero time-like and longitudinal components. The wave equation reduces to the 2-dimensional equation \cite{Melrose2008}.}}
\begin{align}\label{det0}
&\left[k^4+2 k^2 \xi ^2 \left(B l^2+C p^2\right)-4 B C \xi ^4 \left\{(l\cdot p)^2+l^2 p^2\right\}\right]\\
&\times\left[k^4+2 k^2 \xi ^2 \left(B m^2+C n^2\right)-4 B C \xi ^4 \left\{(n\cdot m)^2+m^2 n^2\right\}\right]=0. \nonumber\
\end{align}
One can also verify that the terms multiplying $-4\xi^4BC$ are given by $\{(l\cdot p)^2+l^2 p^2\}=k^2\ k^{(1)}k^{(1)}$, and $\{(n\cdot m)^2+m^2 n^2\}=k^2\ k^{(2)}k^{(2)}=0$, in the equatorial plane. Moreover, a resulting term $-4 B C \xi ^4 k^{(1)}k^{(1)}$ would correspond to a second-order correction to the light cone equation, since it is proportional to the small coefficients $\xi^4$, $B$ and $C$. Hence, a reasonable approximation for the determinant condition is
\begin{equation}\label{maindet}
k^4\left[k^2+2 \xi ^2 \left(B l^2+C p^2\right)\right]\left[k^2+2 \xi ^2 \left(B m^2+C n^2\right)\right]=0\, .
\end{equation}
The first root of this equation \eqref{maindet} corresponds to waves in vacuum with the dispersion relation $k^2=0$. 
The other two roots modify the light cone and give rise to a gravitational birefringence effect. Of course, these two polarizations are corrections to those in the Schwarzschild case, modified by the parameters $\alpha$ and $\beta$ encoded in the functions $B$ and $C$. For both polarizations, we analyze the radial, $k^{(3)}=0$, and the orbital, $k^{(1)}=0$, motions in the equatorial plane $k^{(2)}=0$.

\subsection{The radial polarization}

The radial polarization is derived from the second root of equation \eqref{maindet}, $k^2+2 \xi ^2 \left(B l^2+C p^2\right)=0$. In terms of the Lorentz components $k^{a}$ it reads as
\begin{align}\label{root01}
&-\left[1+2\xi^2B\right]k^{(0)}k^{(0)}+\left[1+2\xi^2(B+C)\right] k^{(1)}k^{(1)}\nonumber\\
&+\left[1+2\xi^2C\right]k^{(3)}k^{(3)}=0\, .
\end{align}
Dividing by $\left[1+2\xi^2B\right]$ and up to first order in $\xi^2$, one can rewrite it as follows
\begin{equation}\label{root1}
-k^{(0)}k^{(0)}+\left[1+2\xi^2C\right] k^{(1)}k^{(1)}
+\left[1-2\xi^2(B-C)\right]k^{(3)}k^{(3)}=0\, .
\end{equation}
Therefore, for radial motion, $k^{(3)}=0$, we obtain
\begin{align}\label{VelRr}
\vert k^{(0)}/k^{(1)}\vert=\left[1+2\xi^2C\right]^{1/2}\approx1+\xi^2C\, .
\end{align}
The photon velocity is modified by $\xi^2C$, which corresponds to a slightly modification of the central object due to the deformation parameter $\alpha$ and the distortion one $\beta$, since $\beta$ and $\alpha$ take small values. However, in the Schwarzschild case, $\alpha=\beta=0$, the function $C=0$ vanishes, and the photon velocity remains unchanged for radial motion.

\begin{figure}
\centering
\includegraphics[width=\hsize]{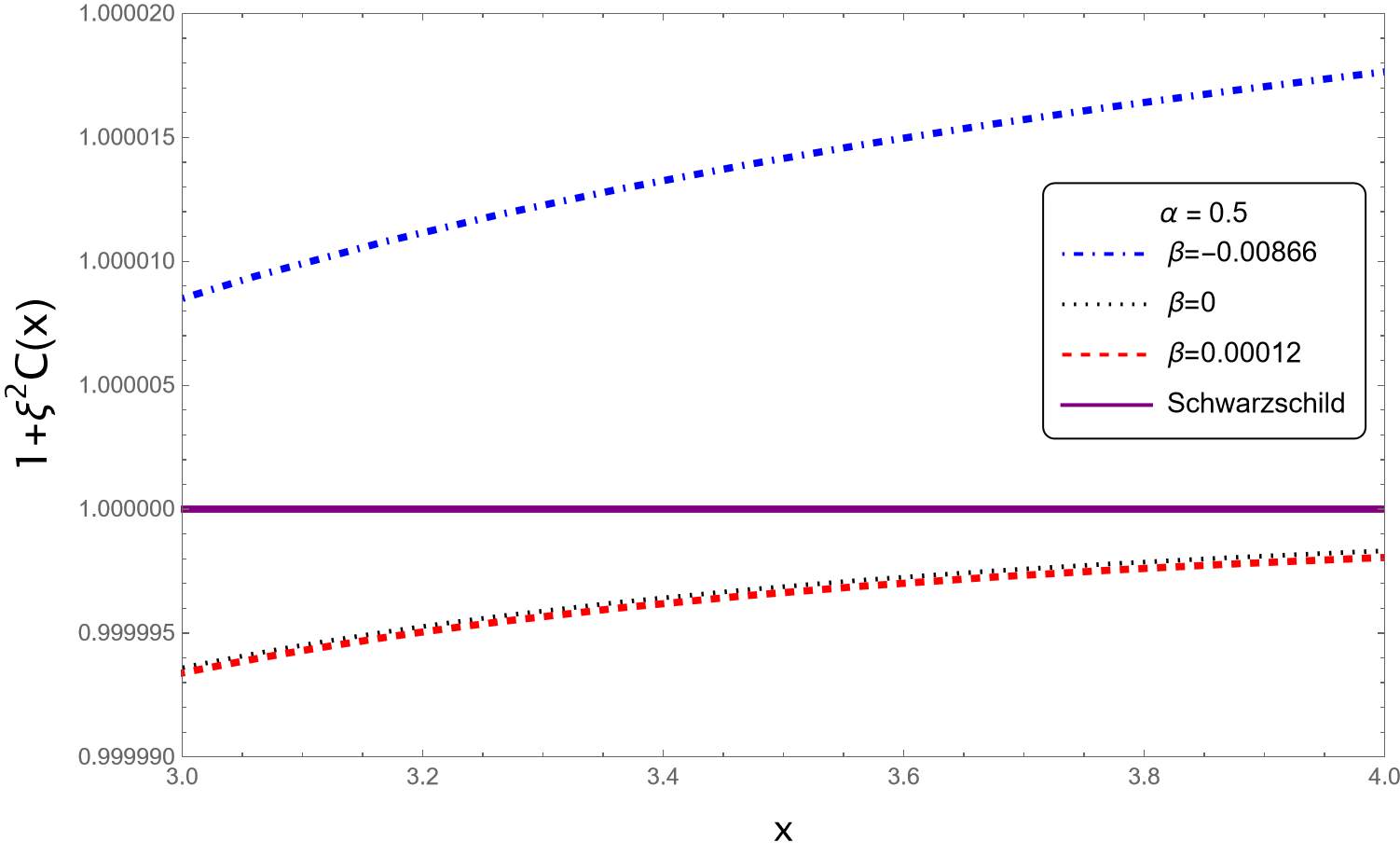}
\includegraphics[width=\hsize]{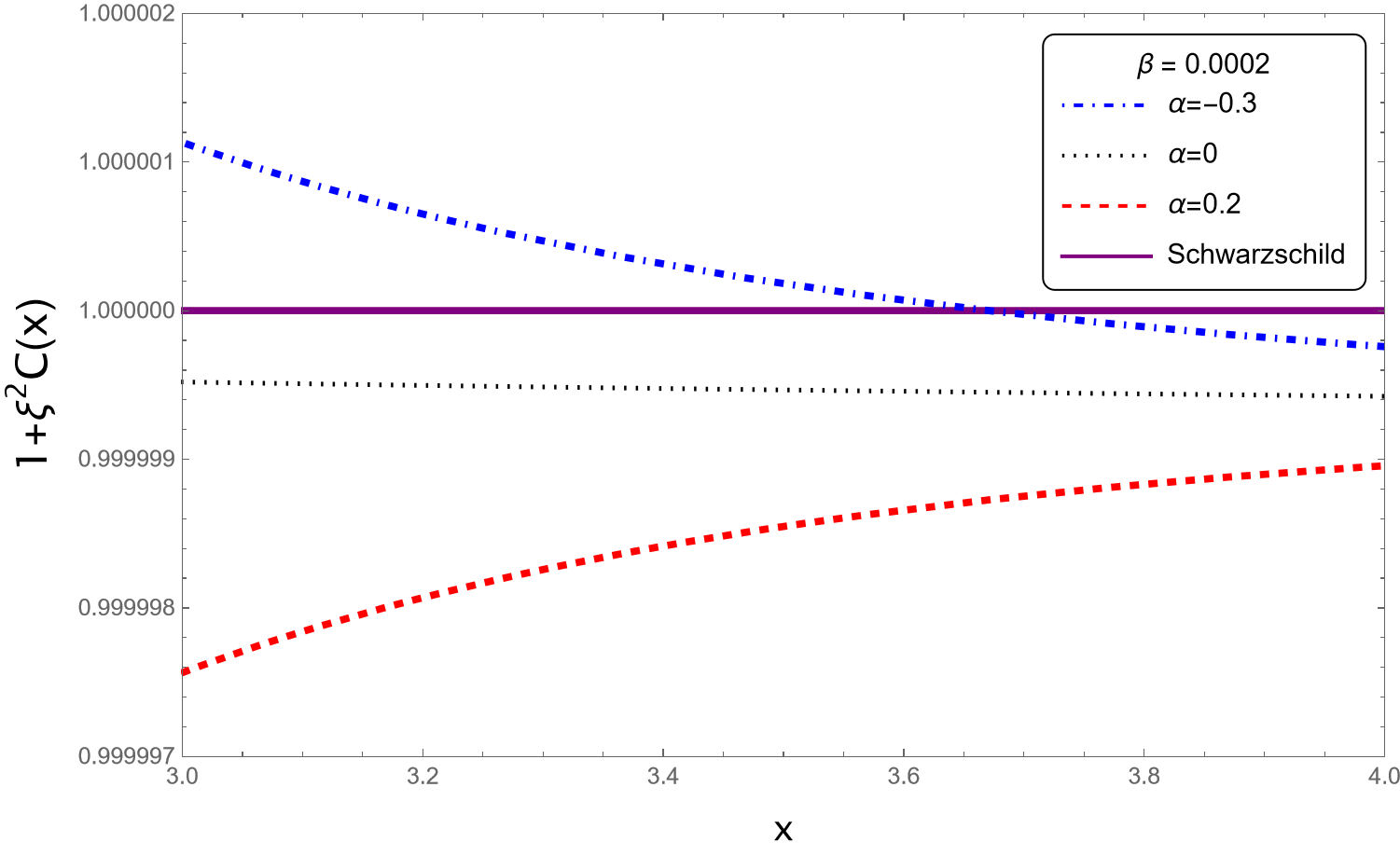}
\caption{\label{figVelRr} The photon velocity for the radial polarization equation and radial motion as a function of the distance $x$, for fixed parameters $M=1$ and $\xi=0.04$, and different values of the parameters $\alpha$ and $\beta$.
}
\end{figure}

This photon velocity relation \eqref{VelRr} can be superluminal, as it is mentioned in section \ref{polar}, depending on the values of the parameters. For instance, in the case $\alpha=0$, the parameter $\beta$ modifies the velocity as follows: from equation \eqref{ABCxq0} the velocity reads as $1-3\xi^2\beta (x-1)/[M^2(x+1)]$, which means that for $x>1$, it is smaller than unity if $\beta$ is positive, and it is superluminal if $\beta$ is negative. Figure \ref{figVelRr} shows an example of this modification.

\begin{figure}
\centering
\includegraphics[width=\hsize]{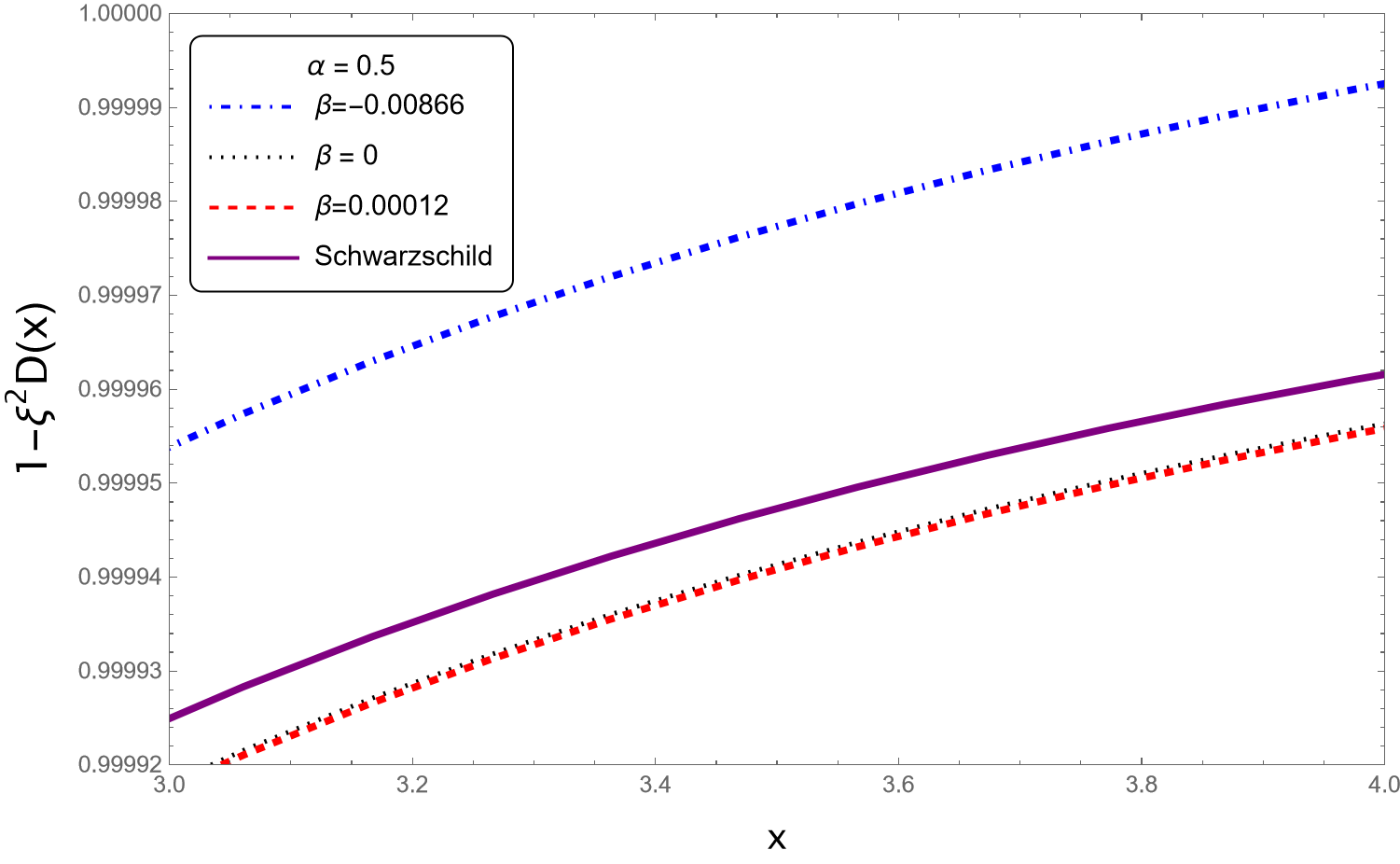}
\includegraphics[width=\hsize]{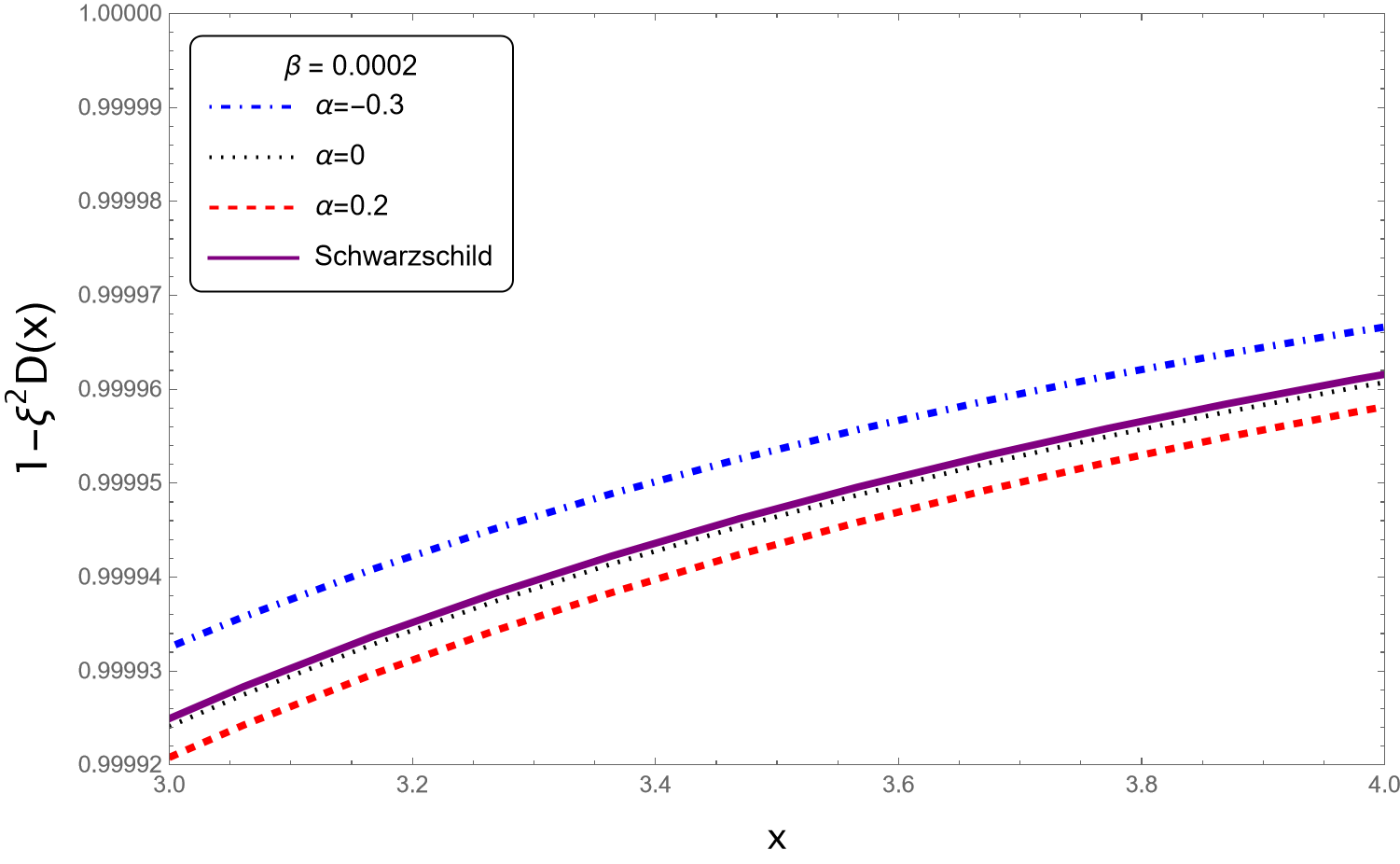}
\caption{\label{figVelRo} The photon velocity for the radial polarization equation and orbital motion as a function of the distance $x$, for fixed parameters $M=1$ and $\xi=0.04$, and different values of the parameters $\alpha$ and $\beta$} (here the function $B-C$ is denoted by $D$).

\end{figure}

On the other hand, for orbital motion, $k^{(1)}=0$, we derive the following
\begin{equation}\label{VelRo}
\vert k^{(0)}/k^{(3)}\vert=\left[1-2\xi^2(B-C)\right]^{1/2}\approx 1-\xi^2(B-C)\, .
\end{equation}
In this case, the polarization has a correction to the Schwarzschild case, which is smaller than unity. Such correction depends strongly on the parameters $\xi$, $\beta$, $\alpha$ and the distance. Figure \ref{figVelRo} shows an example of this photon's velocity for fixed $\alpha$ and $\beta$ values. The effect of $\alpha$ on the photon velocity is more relevant for places closer to the central object. Additionally, like in the previous case, we have a possibility of superluminal velocity for negative values. 

\subsection{The transversal polarization}

\begin{figure}
\centering
\includegraphics[width=\hsize]{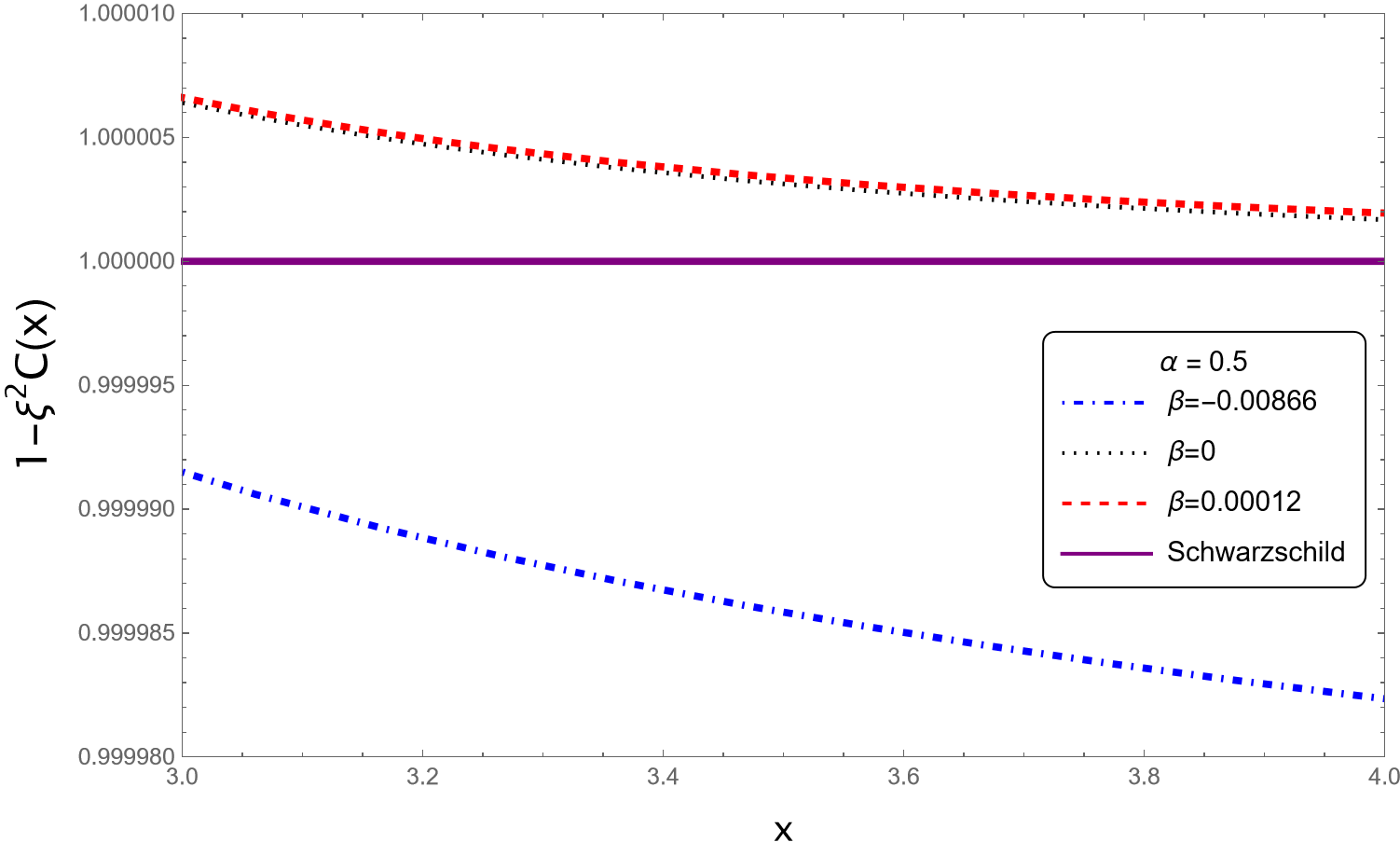}
\includegraphics[width=\hsize]{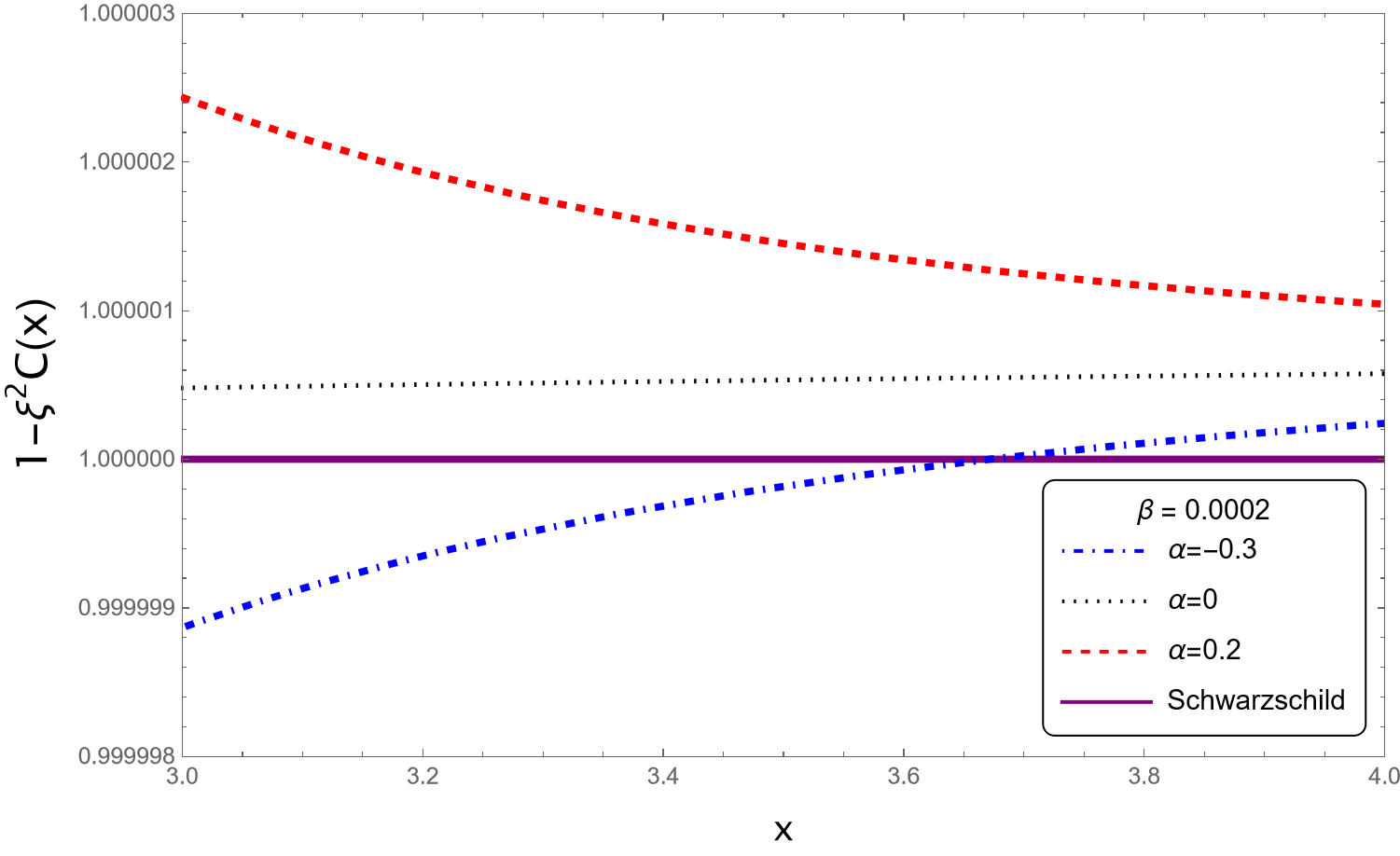}
\caption{\label{figVelTr} The photon velocity for the transversal polarization equation and radial motion as a function of the distance $x$, for fixed parameters $M=1$ and $\xi=0.04$, and for different values of the parameters $\alpha$ and $\beta$.
}
\end{figure}

\begin{figure}
\centering
\includegraphics[width=\hsize]{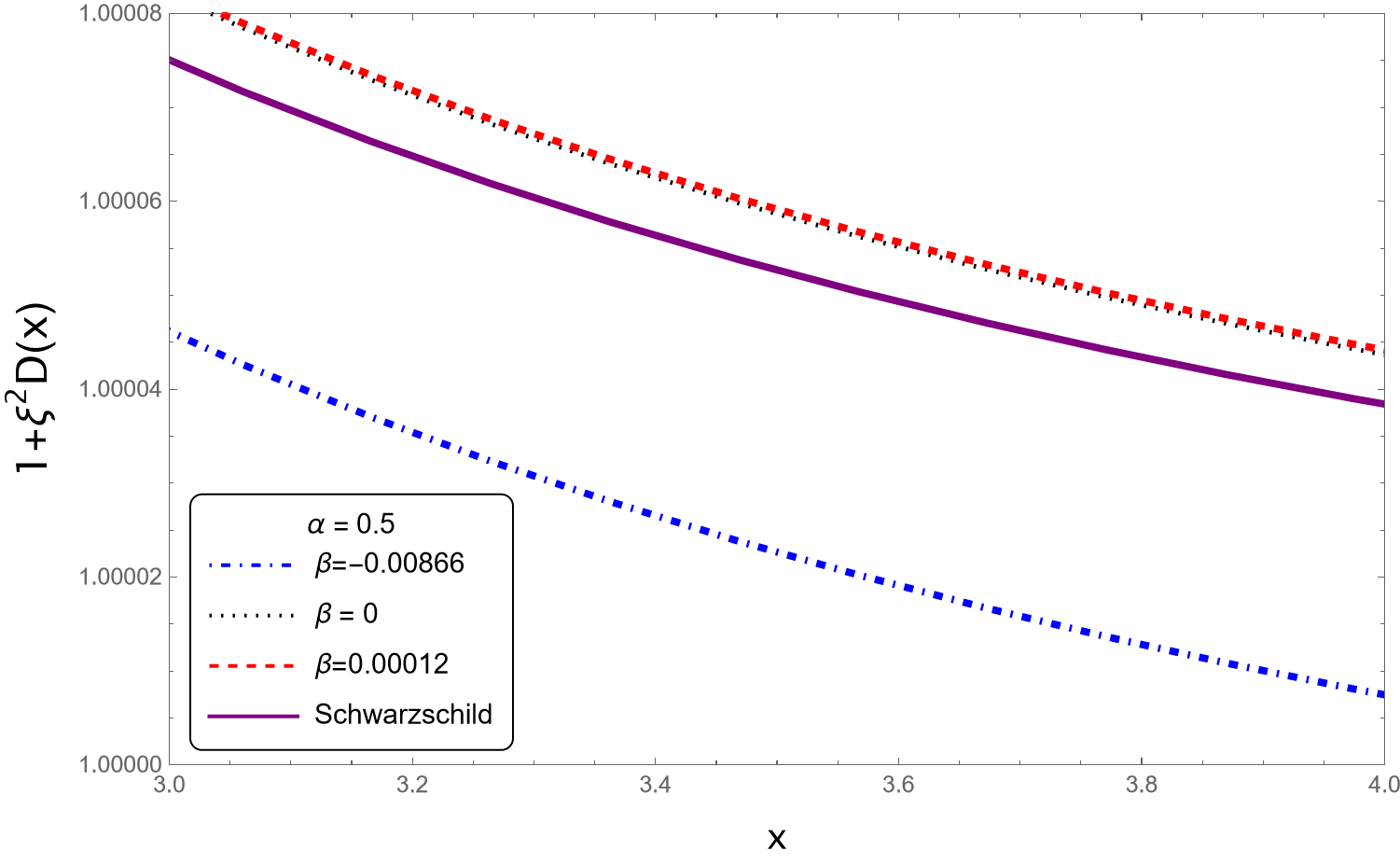}
\includegraphics[width=\hsize]{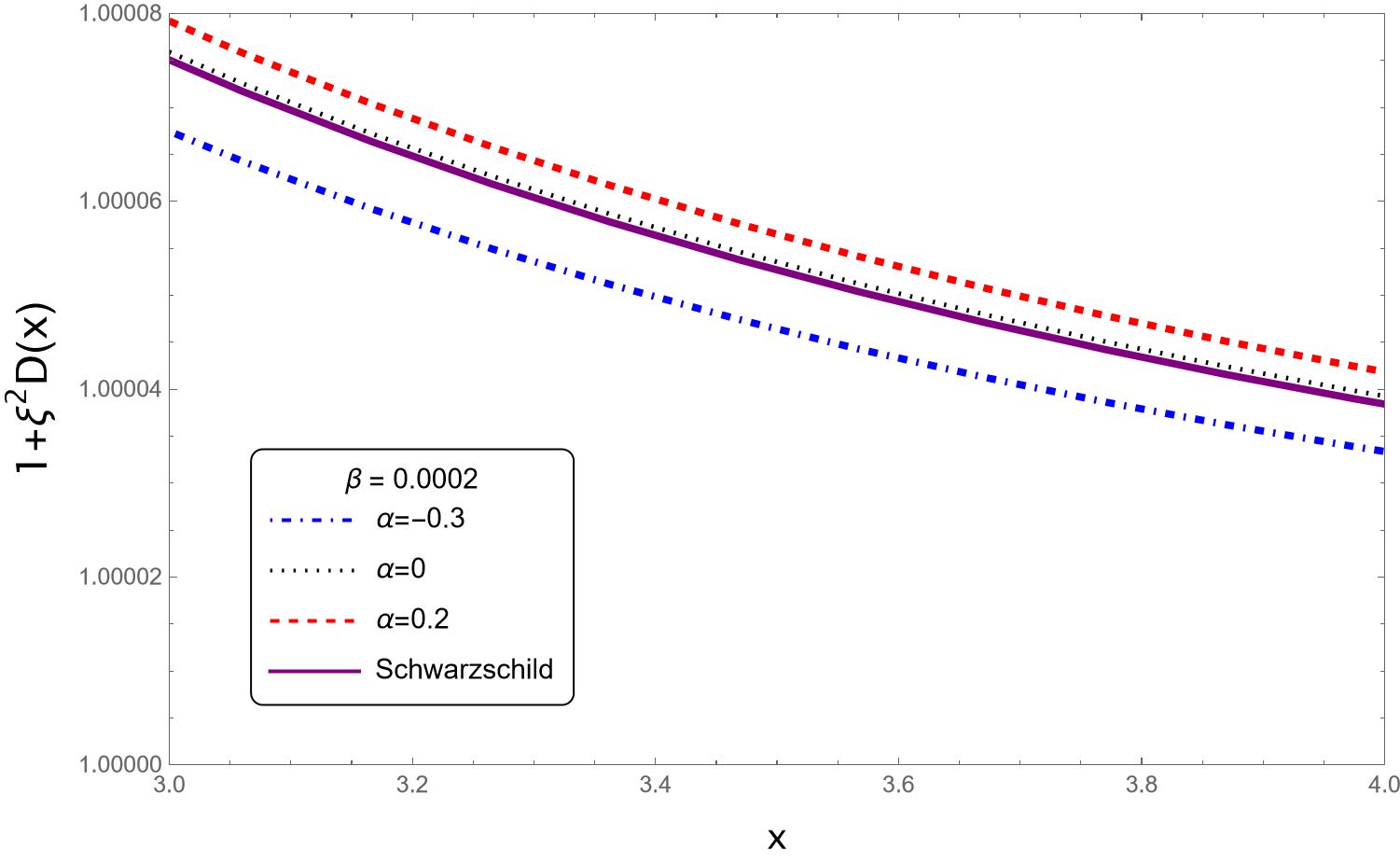}
\caption{\label{figVelTo} The photon velocity for the transversal polarization equation and orbital motion as a function of the distance $x$, for fixed parameters $M=1$ and $\xi=0.04$, and for different values of the parameters $\alpha$ and $\beta$} (here the function $B-C$ is denoted by $D$).

\end{figure}

The transversal polarization corresponds to the last root of the equation \eqref{maindet} given by $k^2+2 \xi ^2 \left(B m^2+C n^2\right)=0$, and in terms of the Lorentz components reads

\begin{equation}\label{root02}
-\left[1+2\xi^2C\right]k^{(0)}k^{(0)}+k^{(1)}k^{(1)}+
\left[1+2\xi^2B\right]k^{(3)}k^{(3)}=0\, .
\end{equation}
Dividing by$ \left[1+2\xi^2C\right]$ and keeping the terms only up to the first order in $\xi^2$, one obtains
\begin{equation}\label{root2}
-k^{(0)}k^{(0)}+\left[1-2\xi^2C\right] k^{(1)}k^{(1)}+
\left[1+2\xi^2(B-C)\right]k^{(3)}k^{(3)}=0\, .
\end{equation}
For the radial motion we now have
\begin{align}
    \vert k^{(0)}/k^{(1)}\vert=1-\xi^2C. \ 
\end{align}
Therefore, there is a small modification on the photon velocity, which can be superluminal for some values of the parameters $\alpha$ and $\beta$. Figure \ref{figVelTr} shows some examples. For the orbital motion we obtain
\begin{equation}\label{velplus}
\vert k^{(0)}/k^{(3)}\vert=1+\xi^2(B-C)\, .
\end{equation}
Hence, the photon with transversal polarization can travel with a superluminal velocity. For instance, for $\alpha=0$, this relation becomes
\begin{equation}
\vert k^{(0)}/k^{(3)}\vert= 1+\frac{3\xi^2}{M^2(x+1)^3}\left[1 +\beta\left(x^3 +2x^2 +4x -1\right)\right]\, .
\end{equation} 
The leading term to the photon velocity is the Schwarzschild one, $1+3\xi^2/[M^2(x+1)^3]$, which is bigger than unity. There is an additional contribution from $\beta$. For $x\geq1$, the correction increases for positive $\beta$, and decreases for negative $\beta$ as seen in Figure \ref{figVelTo}.

To sum up, in the context of electromagnetic birefringence where the velocity of photons depends on their polarization in a background electromagnetic field, the light cone and photon velocity are modified for radial and transversal polarization states during radial motion, although the modifications are very small since they are proportional to the parameters $\alpha$ and $\beta$. However, for orbital motion, the velocity may vary depending on the direction of polarization, which includes a correction to the Schwarzschild case. We can write together both light cone equations, \eqref{root1} and \eqref{root2}, as follows:
\begin{equation}\label{roots}
-k^{(0)}k^{(0)}+\left[1\mp2\xi^2C\right] k^{(1)}k^{(1)}+
\left[1\pm2\xi^2(B-C)\right]k^{(3)}k^{(3)}=0\, ,
\end{equation}
where the upper sign corresponds to the transversal polarization and the lower one to the radial polarization. It is easy to verify that for $C=0$, one retrieves the result for the Schwarzschild case \cite{PhysRevD.22.343} as

\begin{align}
    -k^{(0)}k^{(0)}+k^{(1)}k^{(1)}+\left[1\pm \frac{6\xi^2}{M^2(x+1)^3}
\right]k^{(3)}k^{(3)}=0. 
\end{align}

\section{Gravitational lensing for photons with radial or transversal polarization} \label{lens} 
\begin{figure*}[ht]
\centering
\includegraphics[width=16cm]{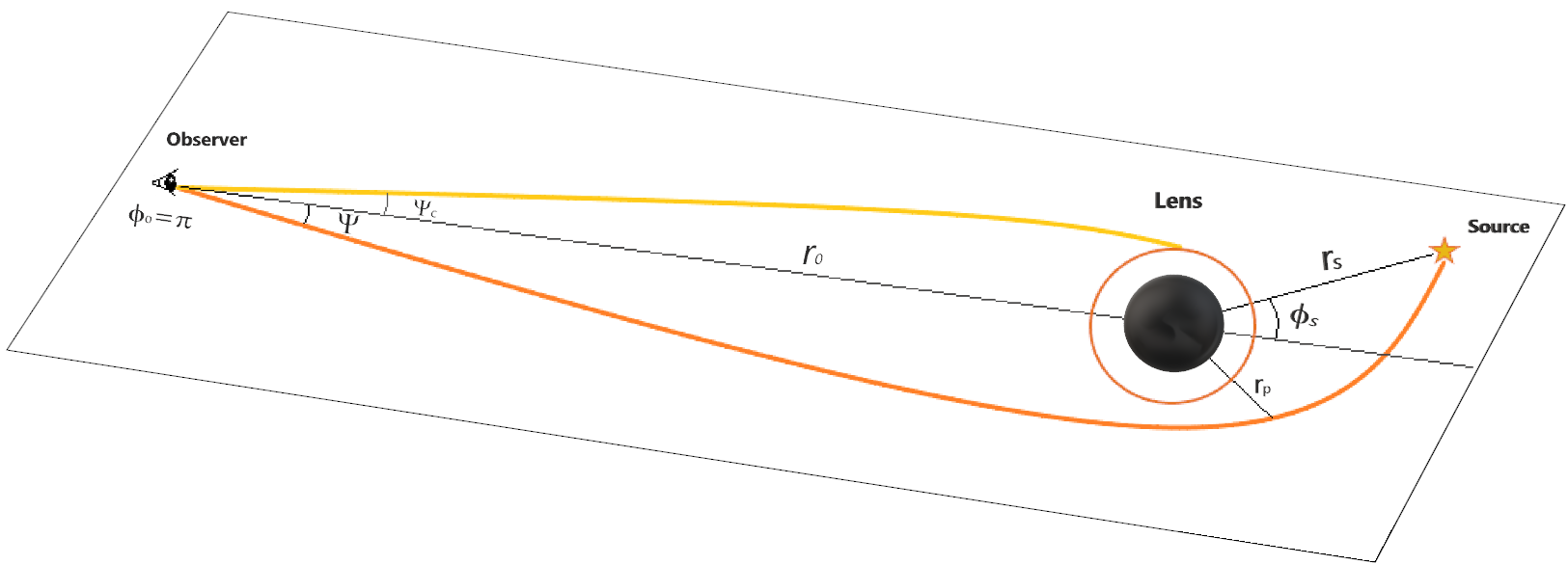}
\caption{\label{figLens} Lens diagram in the equatorial plane. The light ray of a source located at $(r_s,\phi_s)$, reaches the observer's position $(r_0,\phi_0=\pi)$ with incident angle $\Psi$, which determines the angular position of the image. Rays coming from the light ring have critical incident angle $\Psi_c$, and form the edge of the shadow. Only rays with $\Psi\geq\Psi_c$ can be observed. The distance to the point of closest approach is $r_p$.
}
\end{figure*}

Gravitational lensing is a key observational phenomenon that arises due to the bending of light in the presence of a gravitational field. It provides a powerful tool for studying the properties of compact objects and spacetime geometry. For photons, this bending is influenced by the metric of the background spacetime, and additional factors, such as polarization, can further modify the trajectory and observable features. Investigating the effects of polarization on gravitational lensing is particularly important in scenarios involving strong gravitational fields, where higher-order corrections and interactions with spacetime curvature become significant. To illustrate the setup and geometry of the gravitational lensing in the equatorial plane, we present Figure \ref{figLens}, which depicts the trajectory of a light ray originating from a source at $(r_s,\phi_s)$, bending around a compact object, and reaching an observer $(r_0,\phi_0=\pi)$. The figure highlights key parameters such as the incident angle $\Psi$, the angular position of the image, and the critical incident angle $\Psi_c$, which defines the light ring and the edge of the shadow. Only rays with  $\Psi\geq \Psi_c$ are observed, with the point of closest approach denoted by $r_p$. This diagram provides a visual framework for understanding the subsequent analysis of lensing features.

In this section, we revisit the null geodesics in the equatorial plane for photons with radial or transversal polarization, exploring how these polarization states influence lensing phenomena. Specifically, we analyze the impact of polarization on light rings, shadows, and deflection angles, focusing on their observational consequences. To facilitate a direct comparison with previous studies, we present the background metric and equations in Schwarzschild-like coordinates, utilizing the relation (\ref{transf1}). This approach not only provides a clearer connection with prior work but also highlights the distinct effects introduced by polarization in the context of gravitational lensing.

\subsection{Equations of motion}

\begin{figure}[ht]
\centering
\includegraphics[width=\hsize]{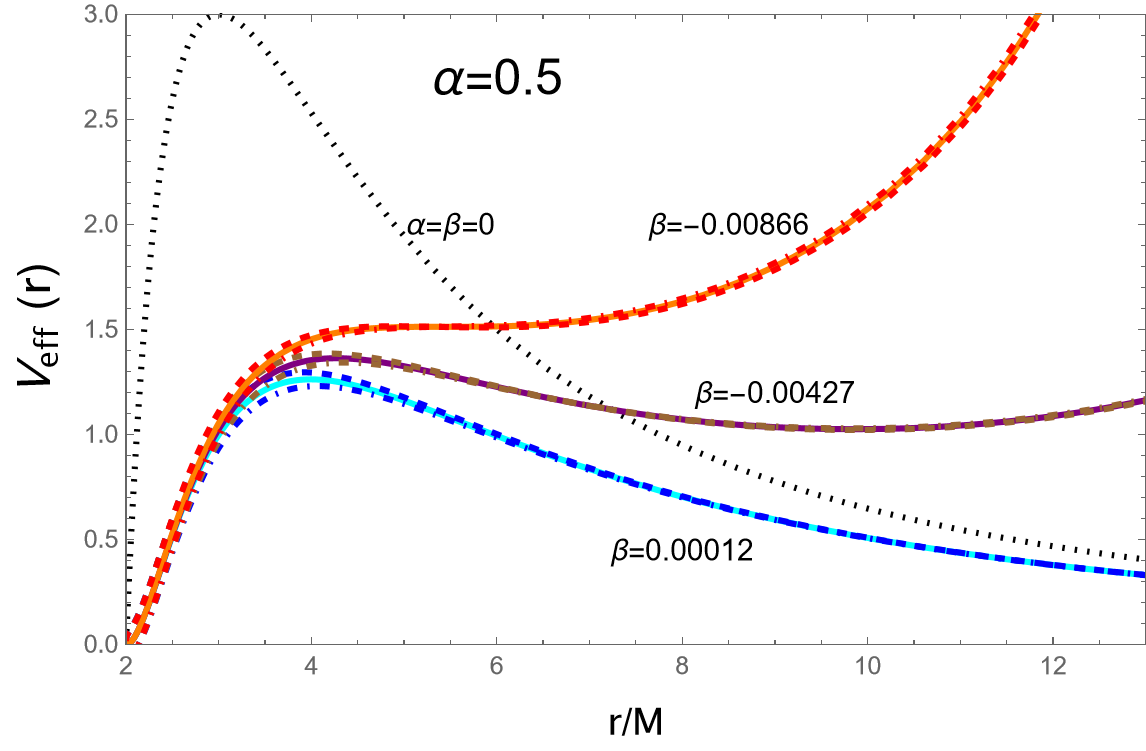}
\includegraphics[width=\hsize]{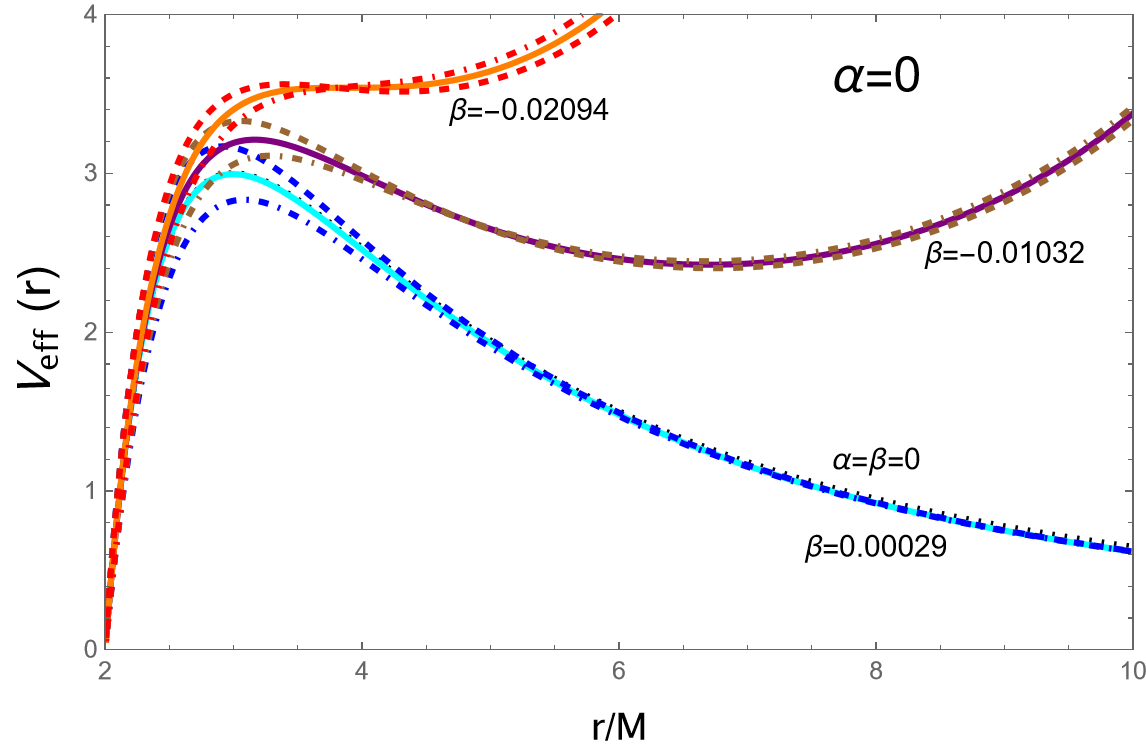}
\includegraphics[width=\hsize]{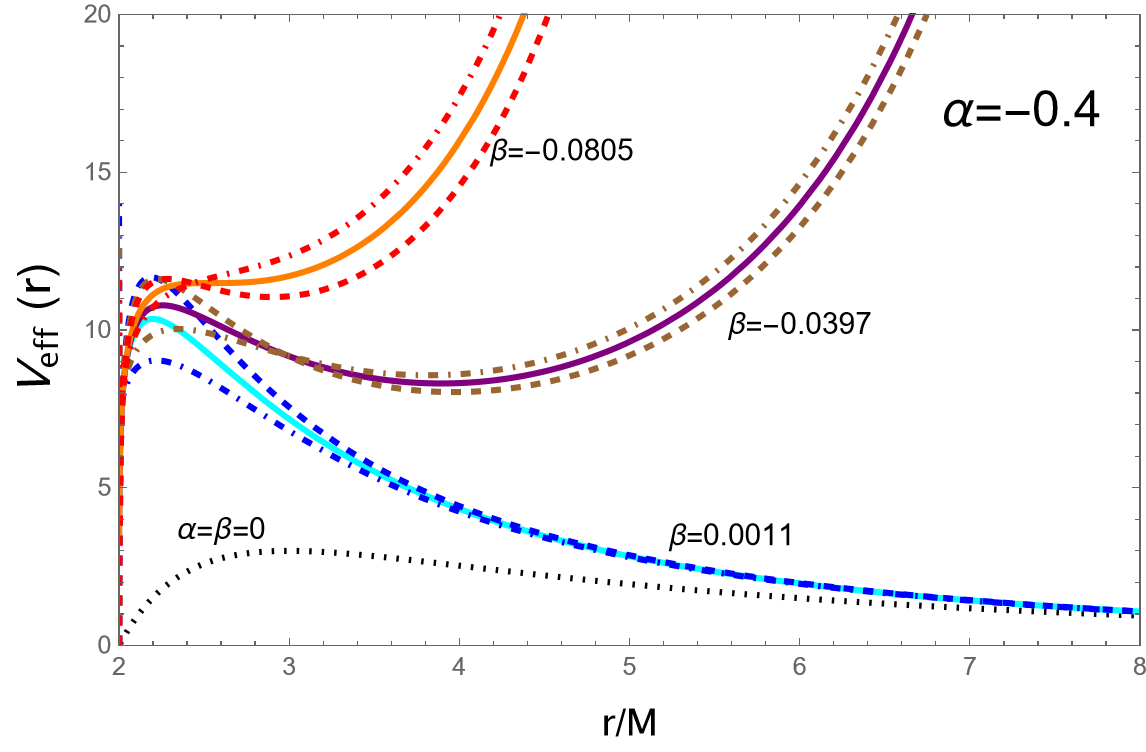}
\caption{\label{figV2} Effective potential as a function of the $r$ coordinate for fixed $M=1$ and $L=9$, and for different values of the parameters $\alpha$ and $\beta$. The dotted line corresponds to the photons in the Schwarzschild metric ($\alpha=\beta=0$) and the bold lines to those in the generalized q-metric ($\xi=0$). The dashed lines correspond to the photons with radial polarization, while the dot-dashed lines to those with transversal one, with $\xi=0.5$.
}
\end{figure}

Polarized photons follow null geodesics of an effective metric $\gamma_{ab}$ defined by the roots (\ref{root1}) and (\ref{root2}) of the light cone condition (\ref{maindet}). To have the effective metric in the coordinate basis, we can express equation \eqref{roots} in the orthonormal basis as $\gamma_{ab}k^{a}k^{b}=0$. The effective metric is then defined as $\gamma_{\mu\nu}=\gamma_{ab}e_{\mu}^{a}e_{\nu}^{b}$, where the tetrads are given by equation \eqref{tetrad}. In the background of the generalized q-metric (\ref{metric}), the photons with transversal/radial polarization will follow null geodesics of the effective metric in the equatorial plane ($\theta=\pi/2$, $\dot{\theta}=\ddot{\theta}=0$) as 
\begin{eqnarray}\label{gamma2}
ds^2&=&-f(r)^{(1+\alpha)}e^{2\psi}dt^2
+r^2f(r)^{-\alpha}e^{-2\psi}\nonumber\\
&&\times\left\{\left[1\mp2\xi^2C(r)\right]\frac{f(r)^{\alpha(2+\alpha)}e^{2\chi}}{\left(1-\frac{M}{r}\right)^{2\alpha(2+\alpha)} }
\frac{dr^2}{r^2f(r)}\right.\nonumber\\
&&\left. \qquad\qquad +\left[1\pm2\xi^2 D(r)\right]d\phi^2\right\}\, ,
\end{eqnarray}
where the functions (\ref{psi}) and (\ref{chi}) read as
\begin{eqnarray}
\psi&=&
-\frac{\beta r^2}{2M^2}f(r)\, ,\label{psiy0}\\
\chi&=&
-\frac{2\beta}{M}\left(1+\alpha\right)(r-M)
+\frac{\beta^2 r^4}{4M^4}f(r)^2\, .\label{chiy0}
\end{eqnarray}
The upper sign in (\ref{gamma2}) corresponds to the transversal polarization, while the lower one to the radial polarization. The function $f(r)=\left(1-\frac{2M}{r}\right)$ is the Schwarzschild metric function,
and $D(r)=B(r)-C(r)$, where $B(r)$ and $C(r)$ are defined in equation \eqref{ABCx}.
In particular, the cases we study below include scenarios where there is only deformation of the central object, i.e., $\alpha \neq 0$ but $\beta = 0$. Then

\begin{eqnarray}
D(r)&=&\frac{M(1+\alpha)}{r^3}\frac{\left[3r^2-3(3+\alpha)Mr
+\left(2\alpha^2+7\alpha+6\right)M^2\right]}
{\left(r^2-3Mr+2M^2\right)}\, \nonumber\\
&&\quad\times\left(1-\frac{2 M}{r}\right)^{\alpha}
\left(\frac{r(r-2M)}{(r-M)^2}\right)^{-\alpha(2+\alpha)}\, ,\label{Dofrb0}
\end{eqnarray}
and
\begin{eqnarray}
C(r)&=&-\frac{M^3}{r^3}\frac{\alpha(1+\alpha)(2+\alpha)}{\left(r^2-3Mr+2 M^2\right)}\, \nonumber\\
&&\quad\times
\left(1-\frac{2M}{r}\right)^{\alpha}\left(\frac{r(r-2M)}{(r-M)^2}\right)^{-\alpha(2+\alpha)}
.\label{Cofrb0}
\end{eqnarray}
While in the case of only distortion, i.e., $\alpha=0$, $\beta\neq0$, they read as
\begin{equation}
D(r)=\frac{3M}{r^3}(1-4\beta)+\frac{3\beta}{M^2r^2}(r^2-Mr+3M^2)\, ,\label{Dofrq0}
\end{equation}
and
\begin{equation}
C(r)=-\frac{3\beta}{M^2}f(r)\, .\label{Cofrq0}
\end{equation}
Additionally, the two conserved quantities are as follows. The energy of the photon, $E=-\gamma_{tt}\dot{t}$, is given by
\begin{equation}
E=f(r)^{(1+\alpha)}e^{2\psi}\ \dot{t}\, ,
\end{equation}
and the angular momentum, $L=\gamma_{\phi\phi}\dot{\phi}$, by
\begin{equation}
L=\left[1\pm2\xi^2D(r)\right]r^2f(r)^{-\alpha} e^{-2\psi}\ \dot{\phi}\, .
\end{equation}
From the normalization condition for null trajectories,
$\gamma_{\mu\nu}\dot{x}^{\mu}\dot{x}^{\nu}=0$, the equation of motion for the $r$-coordinate can be rewritten in terms of an effective potential $V_{\text{eff}}(r)$ described by the equation
\begin{equation}
\left[1\mp2\xi^2C(r)\right]\left(1-\frac{M}{r}\right)^{-2\alpha(2+\alpha)} f(r)^{\alpha(2+\alpha)}e^{2\chi}\ \dot{r}^2 + V_{\text{eff}}(r)=E^2\, ,
\end{equation}
where
\begin{equation}\label{Veff2}
V_{\text{eff}}(r)\equiv \left[1\mp2\xi^2D(r)\right]
\frac{L^2 e^{4\psi}}{r^2}f(r)^{(1+2\alpha)}\, .
\end{equation}
Figure \ref{figV2} shows the effective potential for different values of $\alpha$ and $\beta$. As we see, the shape of the effective potential is strongly influenced by the values of these parameters; however, the effect of $\alpha$ is more vivid and the bigger the value of $\alpha$, the lower the maximum of the potential near $r=3M$. Moreover, what determines if the potential has also minima beyond this maximum, is the value of $\beta$. For example, for some negative values of $\beta$, the potential grows exponentially as $r$ increases. The reason is because the main influence of $\beta$ in the equation (\ref{Veff2}) comes from the exponential $e^{\psi}$. In this case, an outgoing photon would find a turning point and either fall back into the central compact object or reach a bound orbit. Unstable bound orbits are described by the maximum of the potential, while stable ones by the minimum.

The effect of the polarization is more visible near the maximum of the potential, due to the asymptotic behaviour of the function $D(r)$. The potential for photons with transversal polarization is smaller, and then their kinetic energy is bigger, i.e., they travel faster than photons without polarization. For photons with radial polarization, the potential is bigger and the photons slower. Additionally, for the study of the gravitational lensing, we are interested in trajectories that correspond to the case when the effective potential has only a maximum, which would allow the photon to reach a distant observer after meeting a turning point.
The equations of motion simply read as
\begin{eqnarray}
\dot{t}&=& \frac{e^{-2\psi}E}{f^{(1+\alpha)}}
\, ,\label{geot2}\\
\dot{\phi}&=&\left[1\mp2\xi^2D(r)\right]e^{2\psi}f^{\alpha}\frac{L}{r^2}
\, ,\label{geop2}\\
\dot{r}^2&=&\left[1\pm2\xi^2C(r)\right]\left(1-\frac{M}{r}\right)^{2\alpha(2+\alpha)}
f^{-\alpha(2+\alpha)}e^{-2\chi}E^2\times\, \nonumber\\
&&\quad\times\left\{1-\left[1\mp2\xi^2D(r)\right]e^{4\psi}\left(\frac{L}{E}\right)^2
\frac{f^{(1+2\alpha)}}{r^2}\right\}\, . \label{geor2}
\end{eqnarray}
For studying gravitational lensing in the backwards ray-tracing method, it is convenient to rewrite the constants of motion in terms of the celestial coordinates of the light ray as measured by the observer at $r_0$. A photon following a geodesic in the equatorial plane will have an incident angle $\Psi$ with the optical axis defined by \cite{1966MNRAS.131..463S} 

\begin{align}
\cot\Psi:=\sqrt{\frac{\gamma_{rr}}{\gamma_{\phi\phi}}}\frac{dr}{d\phi}.
\end{align}
 Therefore, the following relation is obtained

\begin{eqnarray}\label{LE2}
\left(\frac{L}{E}\right)^2&=&\left[1\pm2\xi^2D(r_0)\right]
\frac{r_0^2\sin^2\Psi}{e^{4\psi(r_0)}f(r_0)^{(1+2\alpha)}}\, .
\end{eqnarray}
Additionally, it is convenient to introduce the inverse radial coordinate
\begin{equation}\label{inv}
u=\frac{1}{r}\, ,
\end{equation}
so, the equation of motion (\ref{geor2}) can be rewritten as
\begin{eqnarray}
\dot{u}^2&=& \left[1\pm2\xi^2C(u)\right](1-Mu)^{2\alpha(2+\alpha)}f(u)^{-\alpha(2+\alpha)} e^{-2\chi(u)}E^2u^4
\times\, \nonumber\\
&&\hspace{-1cm}\times\left\{1-\frac{\left[1\mp2\xi^2D(u)\right]}{\left[1\mp2\xi^2D(u_0)\right]}
\frac{e^{4\psi(u)}}{e^{4\psi(u_0)}}
\frac{u^2f(u)^{(1+2\alpha)}}{u_0^2f(u_0)^{(1+2\alpha)}}
\sin^2\Psi\right\}\, , \label{geou2}
\end{eqnarray}
with $f(u)=1-2Mu$, and $u_0=1/r_0$. 

\subsection{The point of closest approach, the light ring, and the shadow}

We are interested in photons that avoid capture by the strong gravity and subsequently travel to the observer $u_0$. The point of closest approach $u_p$
happens when the light ray reaches a turning point, and it is described by 
the condition $\dot{u}=0$. Then by using equation (\ref{geou2}), we obtain this condition
\begin{eqnarray}
\frac{\left[1\pm2\xi^2D(u_0)\right]\sin^2\Psi}
{e^{4\psi(u_0)}u_0^2f(u_0)^{(1+2\alpha)}}
&=&\frac{\left[1\pm2\xi^2D(u_p)\right]}{e^{4\psi(u_p)}u_p^2f(u_p)^{(1+2\alpha)}}
\, .\label{condpol2}
\end{eqnarray}
Of course, this condition corresponds to the points where the effective potential (\ref{Veff2}) is $V_{\text{eff}}=E^2$. The critical value of $u_p$ is given by the additional condition $\ddot{u}=0$, which corresponds to the condition $V_{\text{eff}}'=0$, i.e.,
to the maxima and minima of the potential (see Figure \ref{figV2} for some examples). However, our focus lies solely on the extrema of the effective potential where photon orbits become unstable, and which compose the light ring at $u_c$. It defines the critical incident angle $\Psi_c$, such that light rays with $\Psi\leq\Psi_c$
are captured by the compact object. By using equation (\ref{geou2}), the condition $\ddot{u}=0$ becomes 
\begin{eqnarray}\label{ucpol2}
\left[1\mp2\xi^2D(u_c)\right]\left[\frac{1}{u_c}-\frac{M(1+2\alpha)}{f(u_c)}
\right.\hspace{1cm}&&\nonumber\\
\left.+\frac{2\beta}{M^2u_c^3}(1-Mu_c)\right] 
\mp \xi^2D'(u_c)&=&0\, .
\end{eqnarray}
For polarized photons i.e., $\xi\neq0$, this equation resists analytical solution. Nevertheless, it can be approached as a correction to the solution derived when $\xi=0$, for which we obtain the cubic polynomial
\begin{equation}\label{ucpol3xi}
M(3+2\alpha)u_{c_0}^3-(1+4\beta)u_{c_0}^2+\frac{6\beta}{M}u_{c_0}-\frac{2\beta}{M^2}=0\, .
\end{equation}
It has either three real roots or one real root and two complex roots, and the real root(s) can be either positive or negative. By using Cardano's method \cite{Nickalls_1993}, the positive real zeros of equation \eqref{ucpol3xi} can be written as
\begin{eqnarray}
u_{c_0}&=&\delta+2\sqrt{{\cal{U}}}\cos
\left[\frac13\arccos {\cal{V}}\right]\, ,\qquad\quad {\cal{V}}>1\, ,\nonumber\\
u_{c_0}&=&\delta+2\sqrt{{\cal{U}}}\cosh
\left[\frac13\text{arccosh} {\cal{V}}\right]\, ,\qquad {\cal{V}}\leq1\, ,\label{soluc}
\end{eqnarray}
where
\begin{eqnarray}
\delta&=&\frac{(1+4\beta)}{3M(3+2\alpha)}\, , \qquad {\cal{U}}=\delta^2-\frac{2\beta}{M^2(3+2\alpha)}
\, ,\nonumber\\
{\cal{V}}&=&\left(\frac{\delta^3-\beta(3\delta-\frac{1}{M^2})}{M^2(3+2\alpha)}\right){\cal{U}}^{-3/2}\, .
\end{eqnarray}
Here, due to the chosen range of parameter values and its definition, ${\cal{U}}$ is always positive. Otherwise, an additional solution involving a $\text{sinh}$ function would be obtained. The solutions in \eqref{soluc} are only valid for $\xi=0$, $\alpha\geq-\frac{19}{18}$, and

\begin{align}
    \beta<\frac{(46+36\alpha)}{32}\left[1-\sqrt{1-\frac{64}{(46+36\alpha)^2}}\right].\label{par_limits}
\end{align}
For instance, the values reported in \cite{universe8030195} satisfy these conditions. Now, to obtain the approximate solutions of the equation (\ref{ucpol2}) for $\xi\neq0$,
we define $u_c=u_{c_0}+\epsilon u_1$, where $u_{c_0}$ is the solution of the equation (\ref{soluc}), and
$\epsilon$ is an infinitesimal parameter. The inverse radial distance of the light ring up to the first order in $\epsilon$ and $\xi^2$ is obtained as

\begin{figure*}[t!]
    \centering
    \begin{tabular}{cc} \includegraphics[width=0.45\hsize]{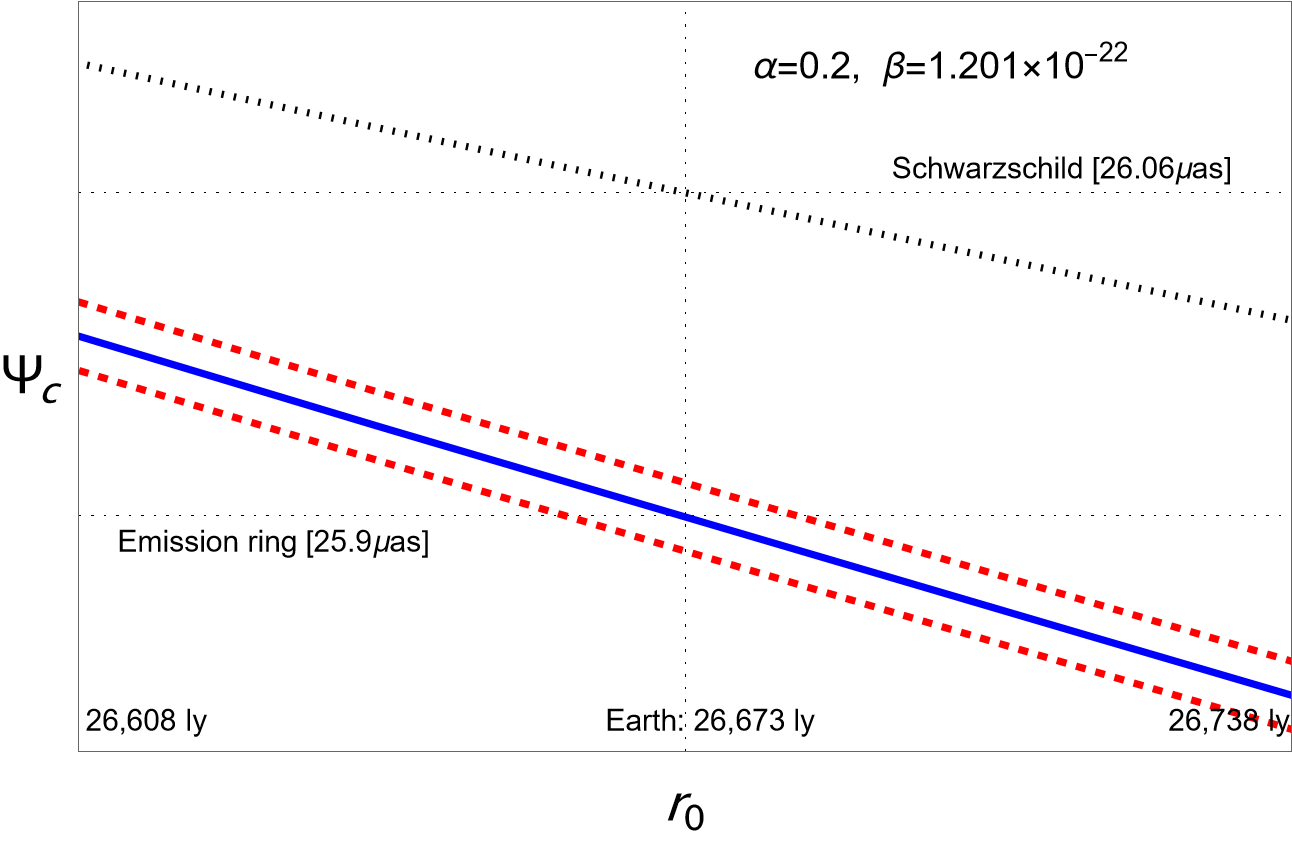}&
\includegraphics[width=0.45\hsize]{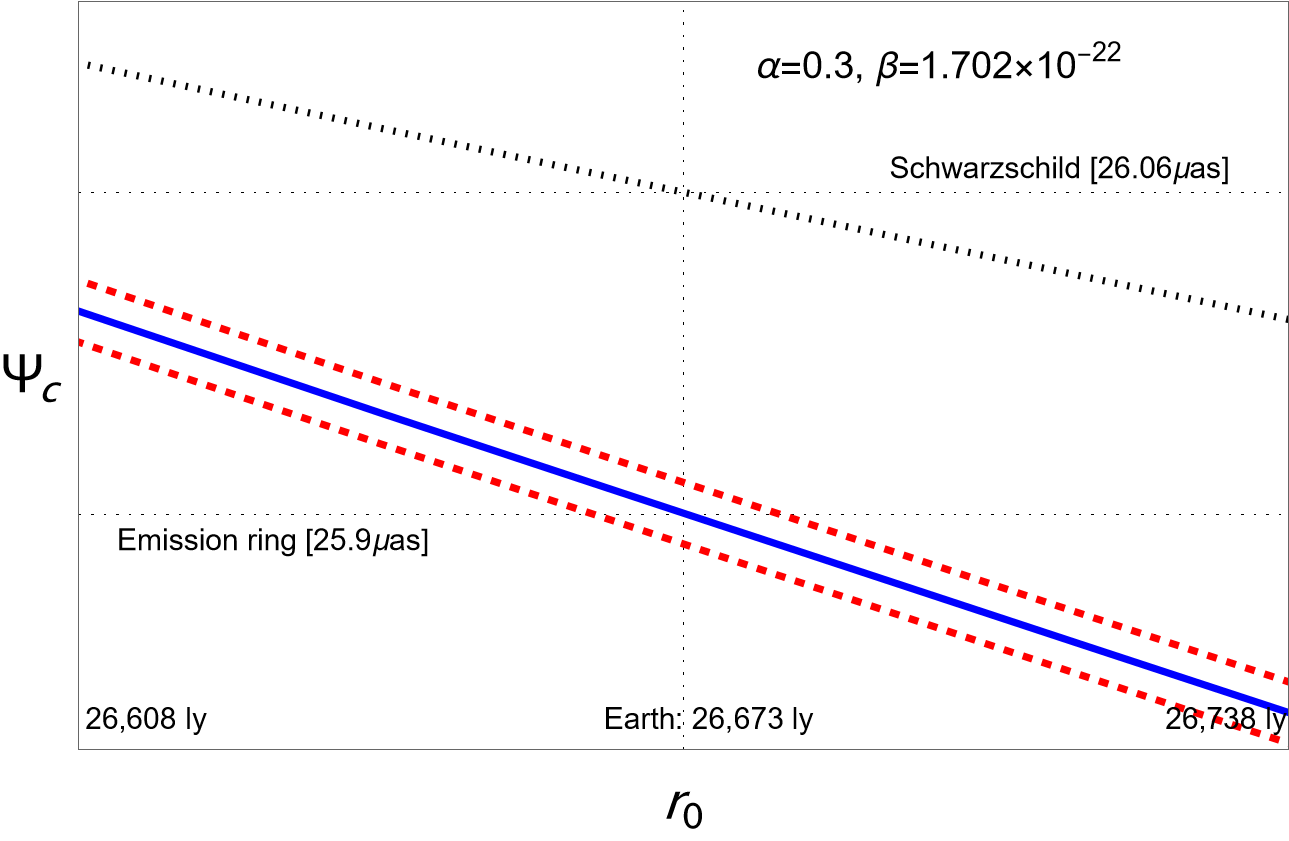}\\
\includegraphics[width=0.45\hsize]{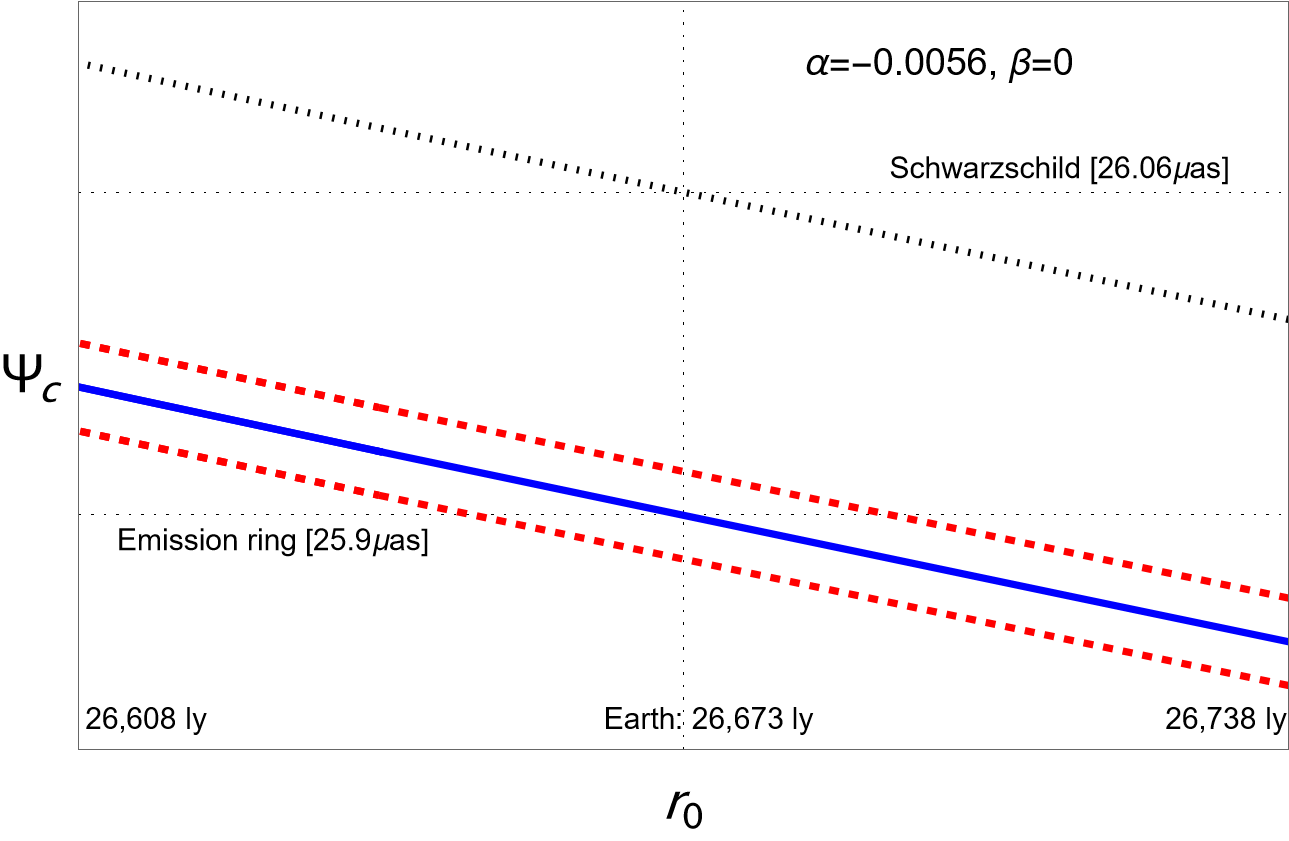}&
\includegraphics[width=0.45\hsize]{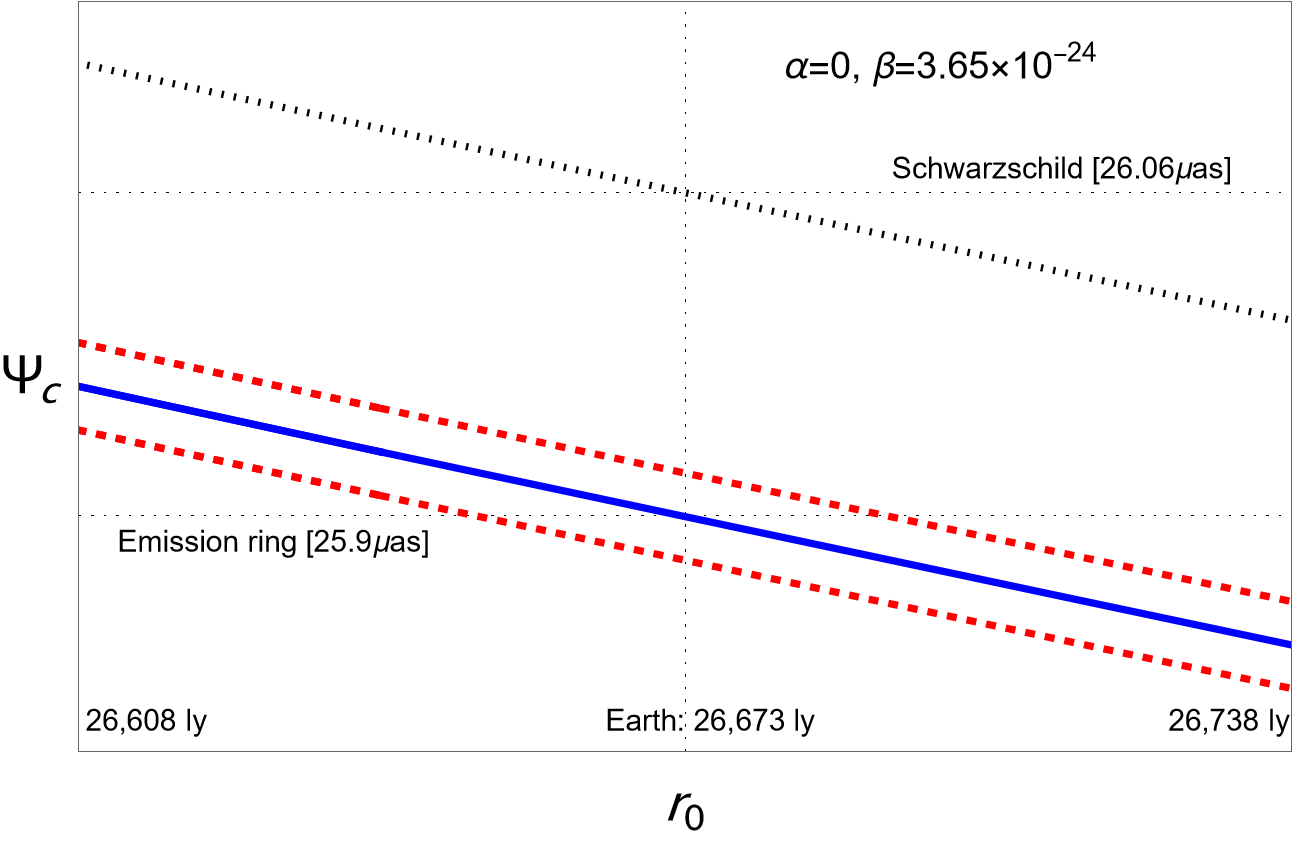}
\end{tabular}
\caption{\label{figSagA} 
The critical incident angle $\Psi_c$ 
as a function of the observer's position $r_0$ for Sagittarius A* with $M=4.154\times 10^{6}M_{\odot}$. The dotted line corresponds to the critical angle predicted by the Schwarzschild metric ($\alpha=\beta=0$). The solid line corresponds to the generalized q-metric with values of $\alpha$ and $\beta$ that match the observed emission ring diameter. 
The upper and lower dashed lines correspond respectively to the transversal and radial polarization, with a fixed value $\xi=0.05$.
}
\end{figure*}

\begin{widetext}
\begin{equation}\label{ucofu2}
u_c=u_{c_0}\left(1\pm\frac{\xi^2u_{c_0}^2(1-2Mu_{c_0})D'(u_{c_0})}{\left[-3M(3+2\alpha)u_{c_0}^2
+2(1+4\beta)u_{c_0}-\frac{6\beta}{M} \mp\xi^2u_{c_0}^2(3-8Mu_{c_0})D'(u_{c_0})\right]}\right)\, .
\end{equation}


\begin{figure}
\centering
\includegraphics[width=8cm]{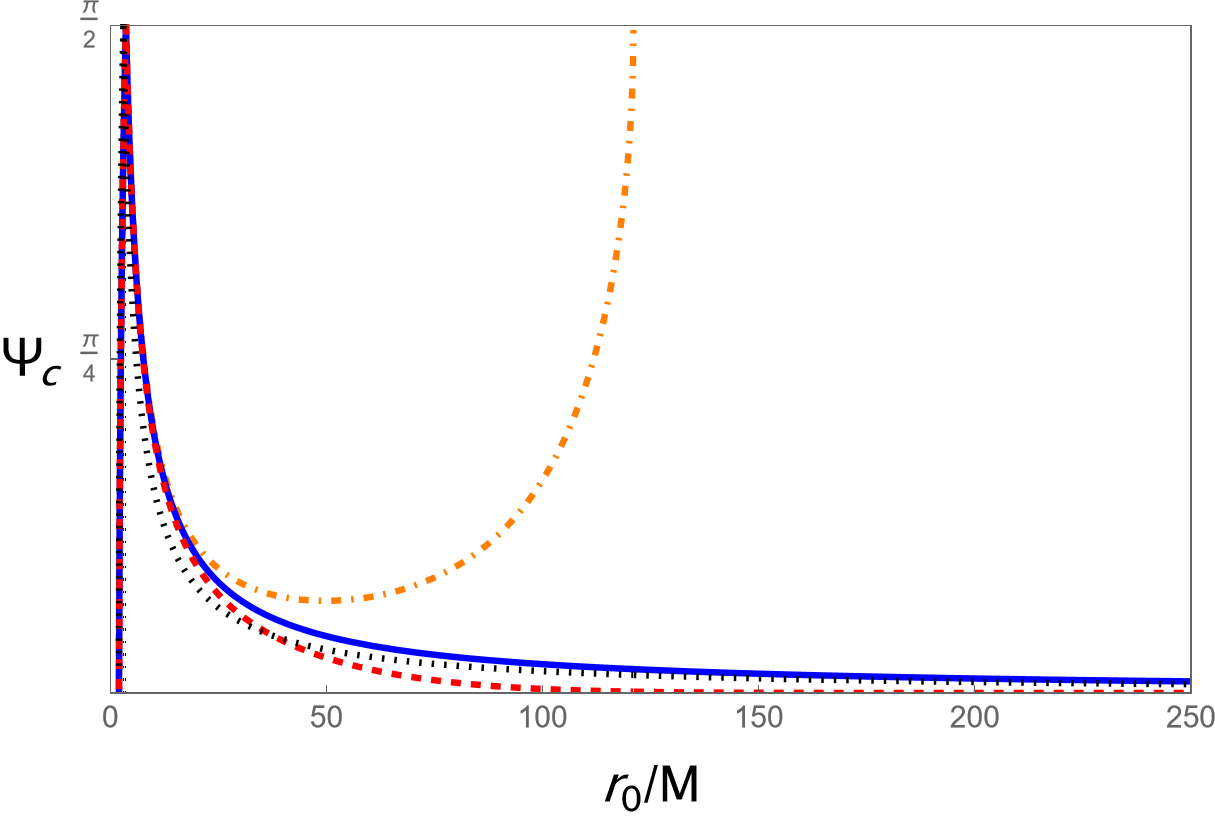}
\includegraphics[width=8cm]{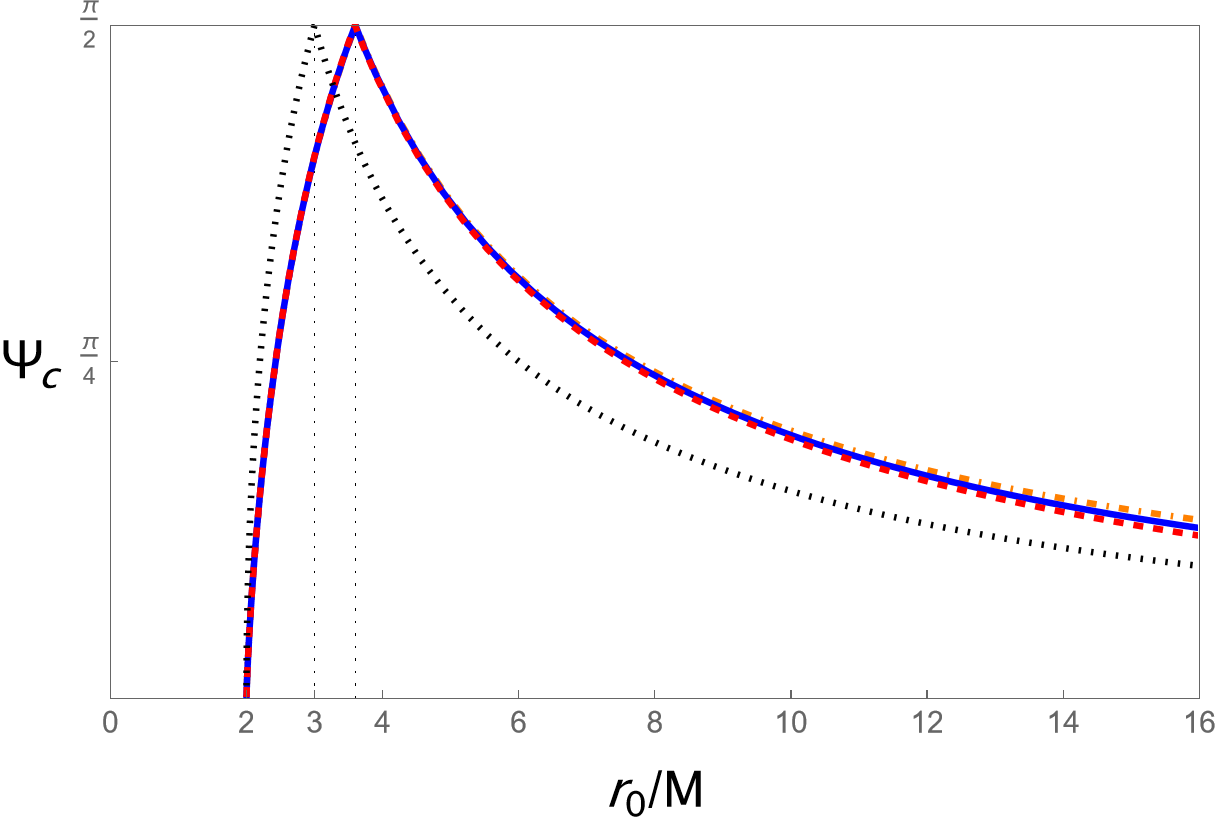}
\caption{\label{figBeha} The critical incident angle $\Psi_c$ 
as a function of the observer's position $r_0$ and for $\xi=0$. On the left, the dotted black curve corresponds
to the Schwarzschild metric ($\alpha=\beta=0$). The other curves correspond to the
generalized q-metric with fixed $\alpha=0.3$ and varying $\beta$.
The case} $\beta=0$ is represented by the solid blue curve, while the the dot-dashed orange curve depicts $\beta=-0.002$, and
$\beta=0.002$ is represented by the dashed red curve.
The value $\Psi_c=\pi/2$ happens when $r_0$ equals the radius of the light ring; for instance,
for Schwarzschild $r_c=3M$, while for $\alpha=0.3$ with $\beta=0$, $r_c=3.6M$.
The latter is illustrated in the figure on the right, which provides a zoomed-in view of the corresponding region from the figure on the left.
\end{figure}


\begin{figure*}
\centering
\includegraphics[width=0.49\hsize]{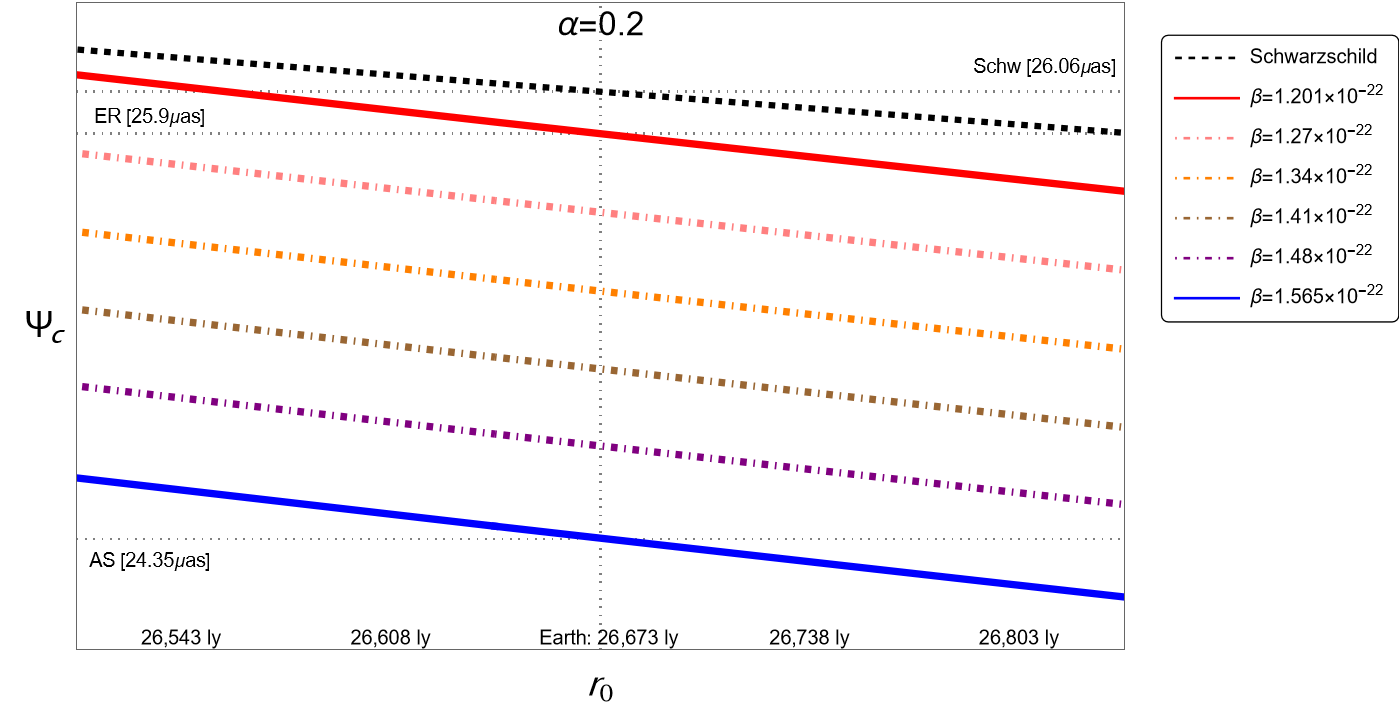}
\includegraphics[width=0.49\hsize]{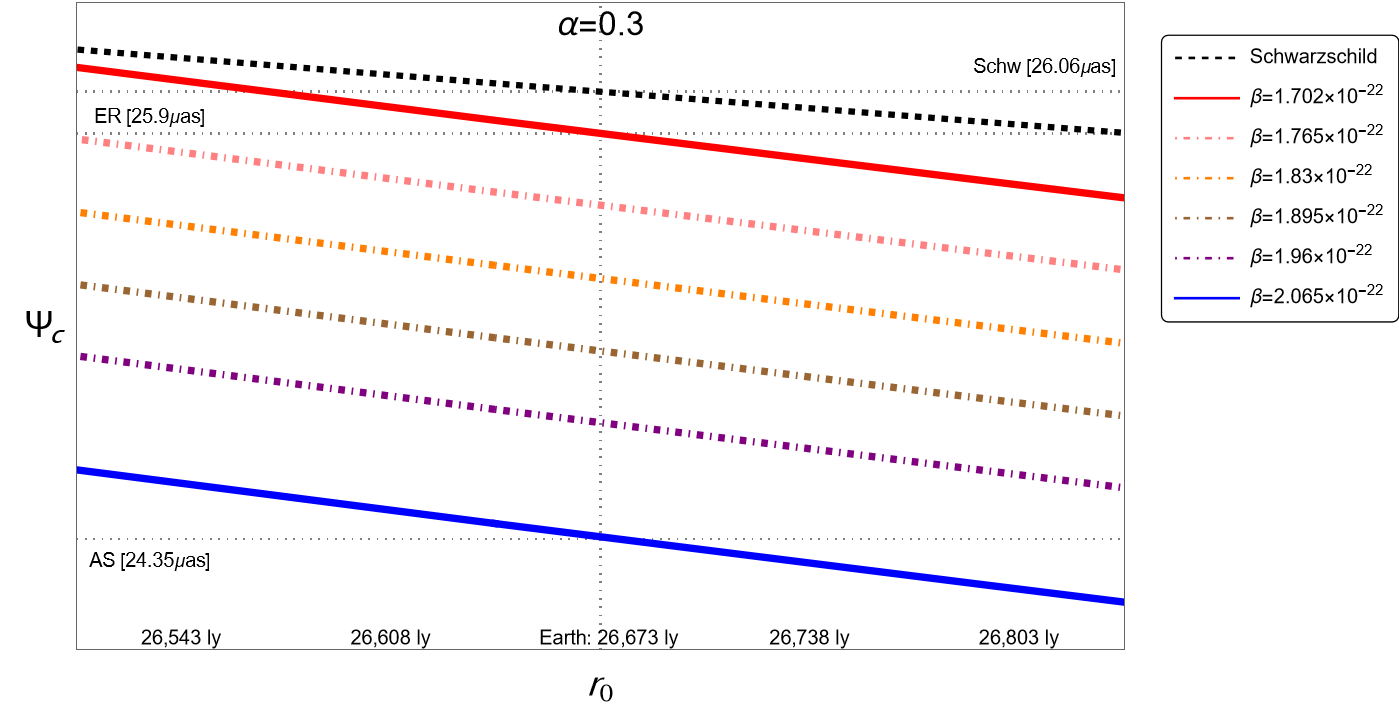}
\caption{\label{figBetas} The critical incident angle $\Psi_c$, 
for $\xi=0$, as a function of the observer's position $r_0$ for Sagittarius A* with $M=4.154\times 10^{6}M_{\odot}$. The black dashed line corresponds to the critical angle predicted by the Schwarzschild metric ($\alpha=\beta=0$). The remaining lines represent the generalized q-metric with a fixed $\alpha$ and varying $\beta$. The upper red solid line corresponds to a $\beta$ value that matches the measured emission ring (ER), while the lower blue solid line corresponds to a $\beta$ value that matches the measured angular shadow (AS). The dot-dashed lines indicate $\beta$ values chosen between these two extremes. As $\beta$ increases, $\Psi_c$ decreases.
}
\end{figure*}

\end{widetext}
The roots of equation \eqref{ucpol2} can also be computed numerically for specific parameter values. To identify the correct root, each numerical solution can be compared with the corresponding approximate solution given in \eqref{ucofu2}. The approximation is reliable, as the difference from the numerical value is minimal, as can be verified. Furthermore, Cardano's method provides constraints on the parameters, such as those outlined in equation \eqref{par_limits}.
By using equation (\ref{condpol2}), the critical incident angle $\Psi_c$ can be written in terms of the observer's inverse radial distance
$u_0$, as 
\begin{eqnarray}\label{Psic2}
\Psi_c&=&\arcsin\sqrt{\frac{\left[1\pm2\xi^2D(u_c)\right]}{\left[1\pm2\xi^2D(u_0)\right]}
\frac{e^{4\psi(u_0)}}{e^{4\psi(u_c)}}\frac{u_0^2f(u_0)^{1+2\alpha}}{u_c^2f(u_c)^{1+2\alpha}}}\, ,\quad
\end{eqnarray}
this is related to the angular radius of the shadow as measured at $u_0$.
Figure \ref{figBeha} shows the critical angle as a function of $r_0=1/u_0$. As we see, positive values of quadrupole moments have the valid range significantly farther from the central object. Conversely, negative values impose severe limitations on the critical incident angle. Nevertheless, the negative sign indicating a prolate distribution of matter has little interest from an astrophysical perspective. In this presentation, our aim is to provide a comprehensive overview of the various possibilities. 
The value $\Psi_c=\pi/2$ occurs when the observer reaches the light ring, i.e., $r_0=r_c$. 
The radius of the shadow $r_{sh}=r_0\tan\Psi_c$ is given by
\begin{equation}\label{shadow0}
r_{sh}=\frac{\sin\Psi_c}{u_0\sqrt{1-\sin^2\Psi_c}}\, .
\end{equation}
For small deflection angles the identity $\tan\Psi_c\approx\sin\Psi_c$ holds.
For a distant observer, the function $D$ becomes $3\beta/M^2$ and the metric function $f$ approaches one. As indicated, this background provides an extended region suitable for observational purposes. In the observations, even though the distance from Earth is significant, it remains finite. For instance, the distance to Sagittarius A* is $r_0\sim 26,673 ly$ \cite{2019A&A...625L..10G}, i.e.,
$r_0\sim 4.113\times10^{10}M$, with $M=4.154\times 10^{6}M_{\odot}$.
The angular diameter of the emission ring is $51.8\pm2.3\mu as$ \cite{2022ApJ...930L..12E},
which corresponds to an angular radius of $\Psi\sim 25.9 \mu as$. The angular shadow diameter
is $48.7\pm7.0\mu as$, and corresponds to $\Psi\sim 24.35 \mu as$. Both values are
smaller than the critical angle for a Shwarzschild black hole in the galactic center
$\Psi_c\sim 26.06 \mu as$. Remarkably, it is possible to determine values for $\beta$ and $\alpha$ such that $\Psi_c$ aligns with the measured angular radius. Various examples illustrating such selections are depicted in Figure \ref{figBetas}. In these representations, we hold $\alpha$ constant while adjusting $\beta$, ensuring that $\Psi_c$ assumes a value between the emission ring and the angular shadow. The plot shows that as $\beta$ increases, there is a corresponding decrease in the value of the critical incident angle $\Psi_c$. This inverse relationship highlights how changes in $\beta$ directly impact the behavior of $\Psi_c$.
In order to analyse the birefringence effect due to the photons with radial and transversal polarizations, in Figure \ref{figSagA} we present the critical incident angle $\Psi_c$ for different values of $\alpha$ and $\beta$. This indicates that, despite the modifications to light propagation being very small (of the order of $\xi^2$), the large distances and small angles involved could make these effects on the shadow of Sagittarius A* detectable in future high-precision observations. For M87*, the angular diameter of the shadow is not as precisely determined as for Sgr A* \cite{2002Natur.419..694S}, but a similar analysis could be feasible with improved measurements. Although here we only display the case for the emission ring, similar plots could be presented for values of $\alpha$ and $\beta$ that match the angular shadow. 

\subsection{The lens equation}

In the context of the backwards ray-tracing method, we examine the photon's trajectory starting from the observer's position at $(r_0, \phi_0=\pi)$ and ending at the source's position at $(r_s, \phi_s)$, (as depicted in the lens diagram in Figure \ref{figLens}). When an observer measures an image angle, $\Psi$, and determines the source location by solving an equation, this equation is referred to as the \textit{lens equation}. The exact lens equation is defined by the geodesic motion of photons \cite{PhysRevD.59.124001, PhysRevD.61.064021}.

From equations (\ref{geop2}) and (\ref{geou2}), and expressed in terms of the inverse radial distance $u=1/r$, one obtains
\begin{widetext}
\begin{eqnarray}\label{intp0}
\phi_s=\pi-\int_{u_0}^{u_s} \frac{\left[1\mp2\xi^2\{2D(u)-C(u)\}\right] e^{2\psi(u)+\chi(u)}f(u)^{(2\alpha+\alpha^2/2)}\left(\frac{L}{E}\right)\ du}
{(1-Mu)^{\alpha(2+\alpha)}\sqrt{1-\left[1\mp2\xi^2D(u)\right]\left(\frac{L}{E}\right)^2
e^{4\psi(u)}u^2f(u)^{(1+2\alpha)}}}\, .
\end{eqnarray}
The trajectories can be divided into two segments based on the direction of the light ray:
\begin{enumerate}
    \item \textbf{First Segment:} The light ray travels inward (\(\dot{r} < 0\)) from the observer at \(r_0\) to the point of closest approach at \(r_p\). In terms of the inverse radial coordinate \(u = 1/r\), this corresponds to the integration from \(u_0 = 1/r_0\) to \(u_p = 1/r_p\), where the inverse radial distance increases (\(\dot{u} > 0\)).
        \item \textbf{Second Segment:} The light ray moves outward (\(\dot{r} > 0\)) from the point of closest approach at \(r_p\) to the source at \(r_s\). In this case, the integration proceeds from \(u_p = 1/r_p\) to \(u_s = 1/r_s\), but the inverse radial distance decreases (\(\dot{u} < 0\)).
\end{enumerate}
As a result, there is a change in the sign of \(\dot{u}\) between the two segments. This change can be handled by treating the integration over the second segment as having swapped limits. This is consistent with the physical system modeled, where \(u_0 = 1/r_0\) corresponds to the observer's position, \(u_p = 1/r_p\) to the closest approach of the photon, and \(u_s = 1/r_s\) to the source's position. These limits are consistent with the equation of motion describing the light ray trajectory in the context of gravitational lensing. Hence, equation (\ref{intp0}) can be written as

\begin{eqnarray}\label{intp1}
\phi_s(u_0,u_s,u_p)&=&\pi-\left(\int_{u_0}^{u_p}
+\int_{u_s}^{u_p} \right)
\frac{\left[1\mp2\xi^2\{2D(u)-C(u)\}\right] e^{2\psi(u)+\chi(u)}f(u)^{(2\alpha+\alpha^2/2)}\left(\frac{L}{E}\right)\ du}
{(1-Mu)^{\alpha(2+\alpha)}\sqrt{1-\left[1\mp2\xi^2D(u)\right]\left(\frac{L}{E}\right)^2
e^{4\psi(u)}u^2f(u)^{(1+2\alpha)}}}\, .
\end{eqnarray}
\end{widetext}
This equation applies to general positions of the source and observer. For the case where the observer is closer to the source, $r_s \geq r_0$, the trajectory of an outgoing light ray from $u_p$ to $u_s$ can be divided into two segments: from $u_p$ to $u_0$ and then from $u_0$ to $u_s$. This approach is commonly applied to study the deflection of light from a distant star due to the gravitational field of the Sun. Typically, this is analyzed using thin-lens approaches \cite{2000PhRvD..62h4003V,2002PhRvD..66j3001B}, which assume that both the source and the observer are located at infinity ($u_s=u_0=0$).  While thin-lens methods are limited, they provide a good approximation for certain specific systems. In contrast, equation (\ref{intp1}) enables the exploration of more general and intriguing scenarios. For instance, it allows the study of cases where the source is closer to the compact object than the observer i.e., $r_s < r_0$. This is crucial for studies of star clusters \cite{2002Natur.419..694S,2021arXiv210213000G,1998ApJ...509..678G}, as well as for examining sources in accretion processes that contribute to the formation of shadows \cite{2022ApJ...930L..12E,2019ApJ...875L...1E}. Additionally, subtle effects such as birefringence due to polarization corrections in light propagation, or variations in metric parameters, become more significant in the strong field regime near the compact object. 


Additionally, $\phi_s=\phi_s(u_0,u_s,\Psi)$ depends on $\Psi$ via equation (\ref{condpol2}). If $u_s$ and $u_0$ are known, equation (\ref{intp1}) can be interpreted as the lens equation, since it determines the angular position of the source from a measured angular position of its image in the celestial sphere of the observer \cite{PhysRevD.61.064021}. Values outside the range $-\pi < \phi_s < \pi$ correspond to multiple circlings around the lens, resulting in multiple images of the same source appearing at different image angles $\Psi$. Additionally, Einstein rings are observed when the source is aligned with the lens and the observer, i.e., when $\phi_s = ..., -4\pi, -2\pi, 0, 2\pi, 4\pi, ...$ . 

\begin{figure*}
\centering
\begin{tabular}{cc} 
\includegraphics[width=0.45\hsize]{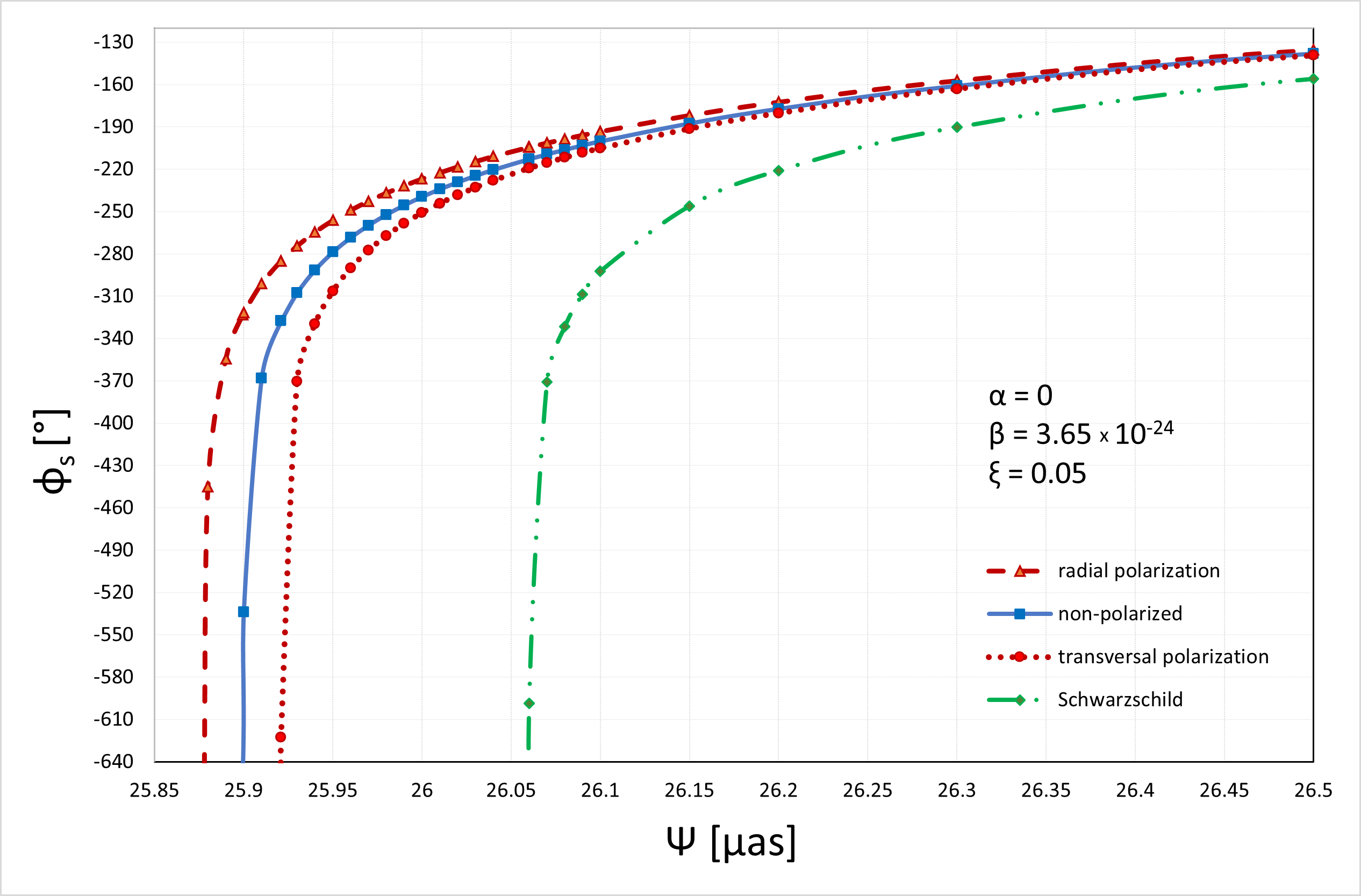} &
\includegraphics[width=0.45\hsize]{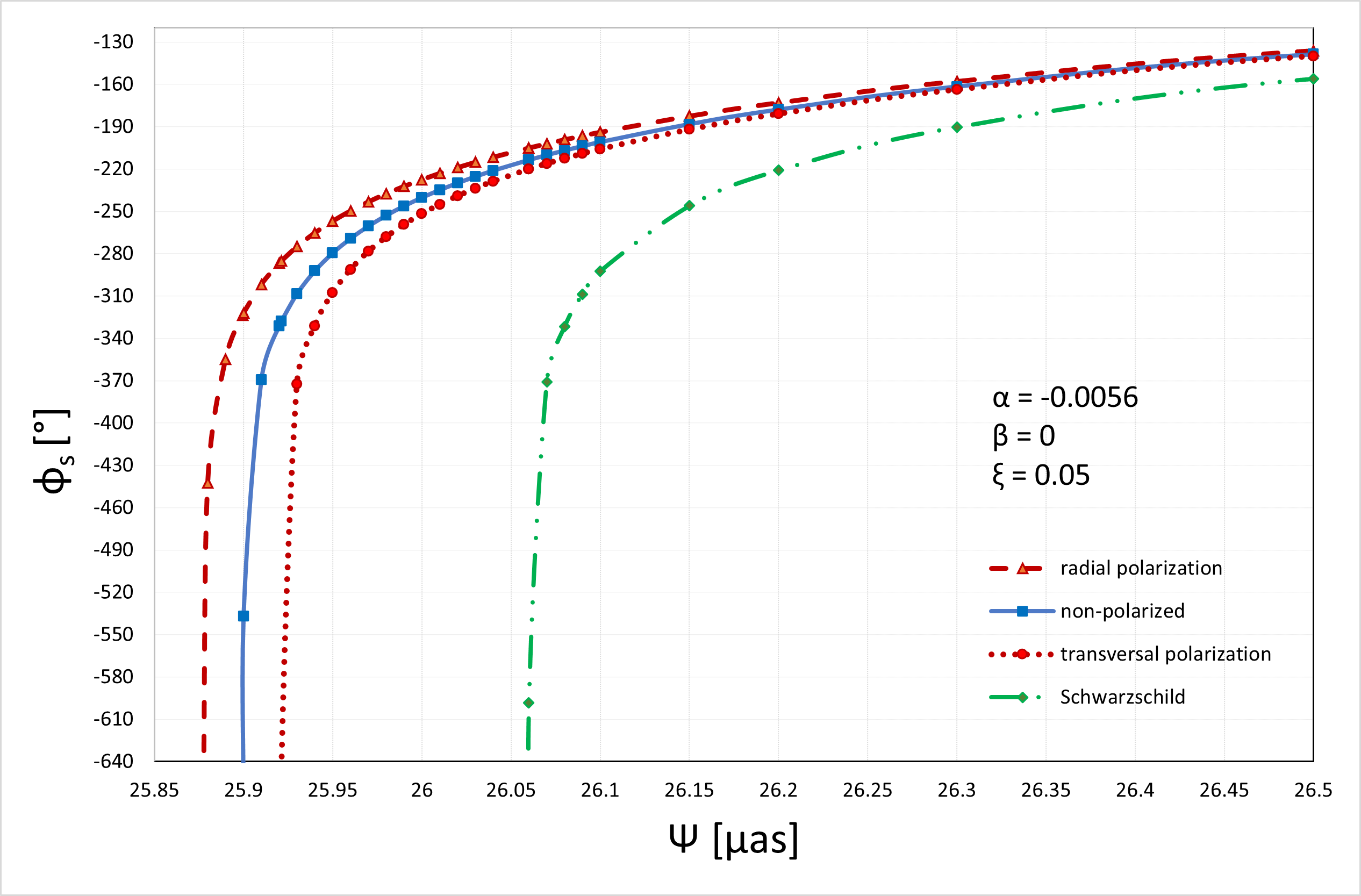} \\ 
\end{tabular}
\caption{\label{figPhi} The angular position $\phi_s$ of a source at $r_s=6M$, as a function of the image angle $\Psi$ for Sagittarius A* with $M=4.154\times 10^{6}M_{\odot}$. The dot-dashed line corresponds to the Schwarzschild case, while the bold line corresponds to the generalized q-metric with values of $\alpha$ and $\beta$ that match the observed angular radius of the emission ring.}
\end{figure*}

\begin{figure*}[ht]
\centering
\includegraphics[width=0.45\hsize]{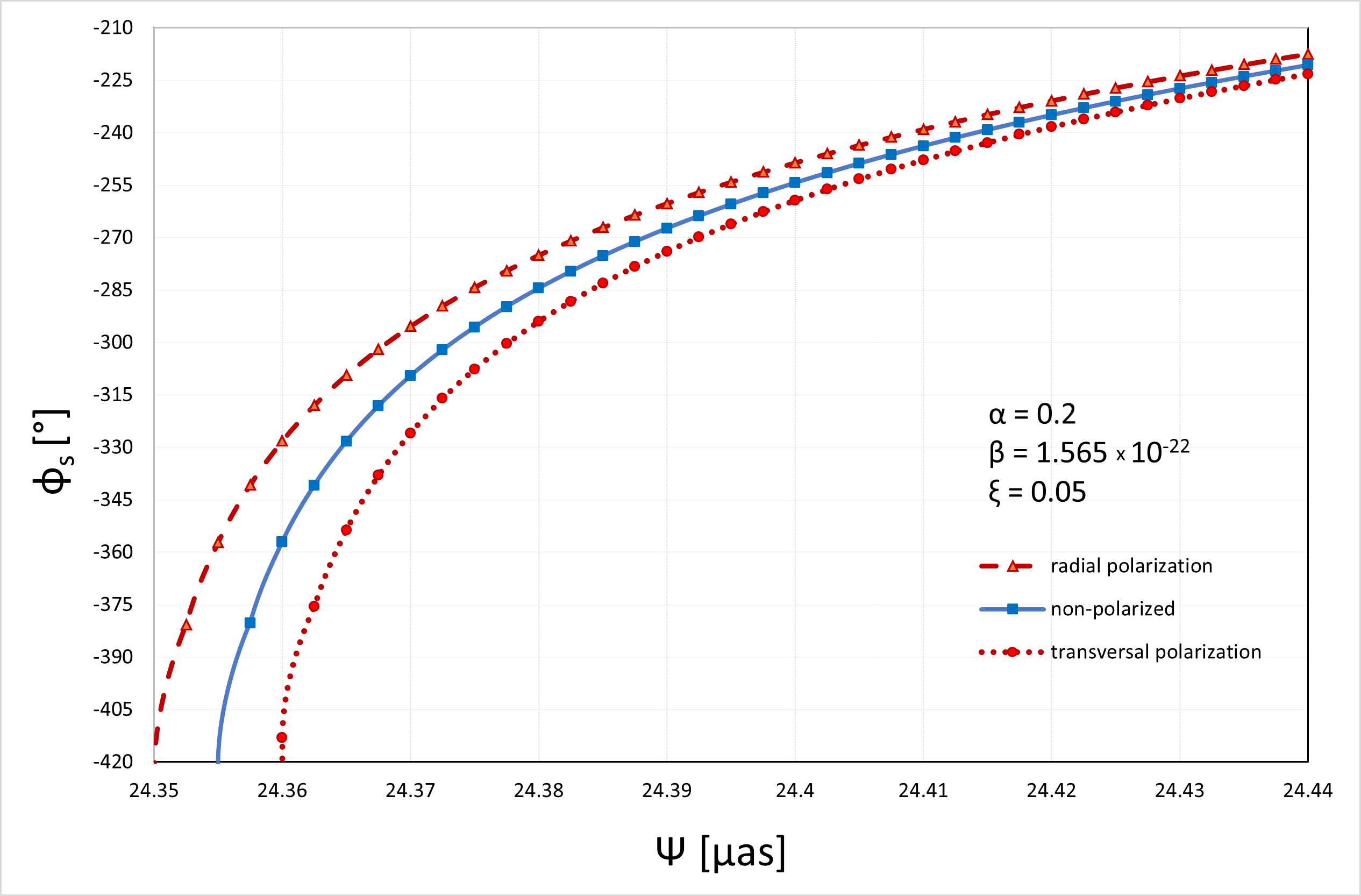}
\includegraphics[width=0.45\hsize]{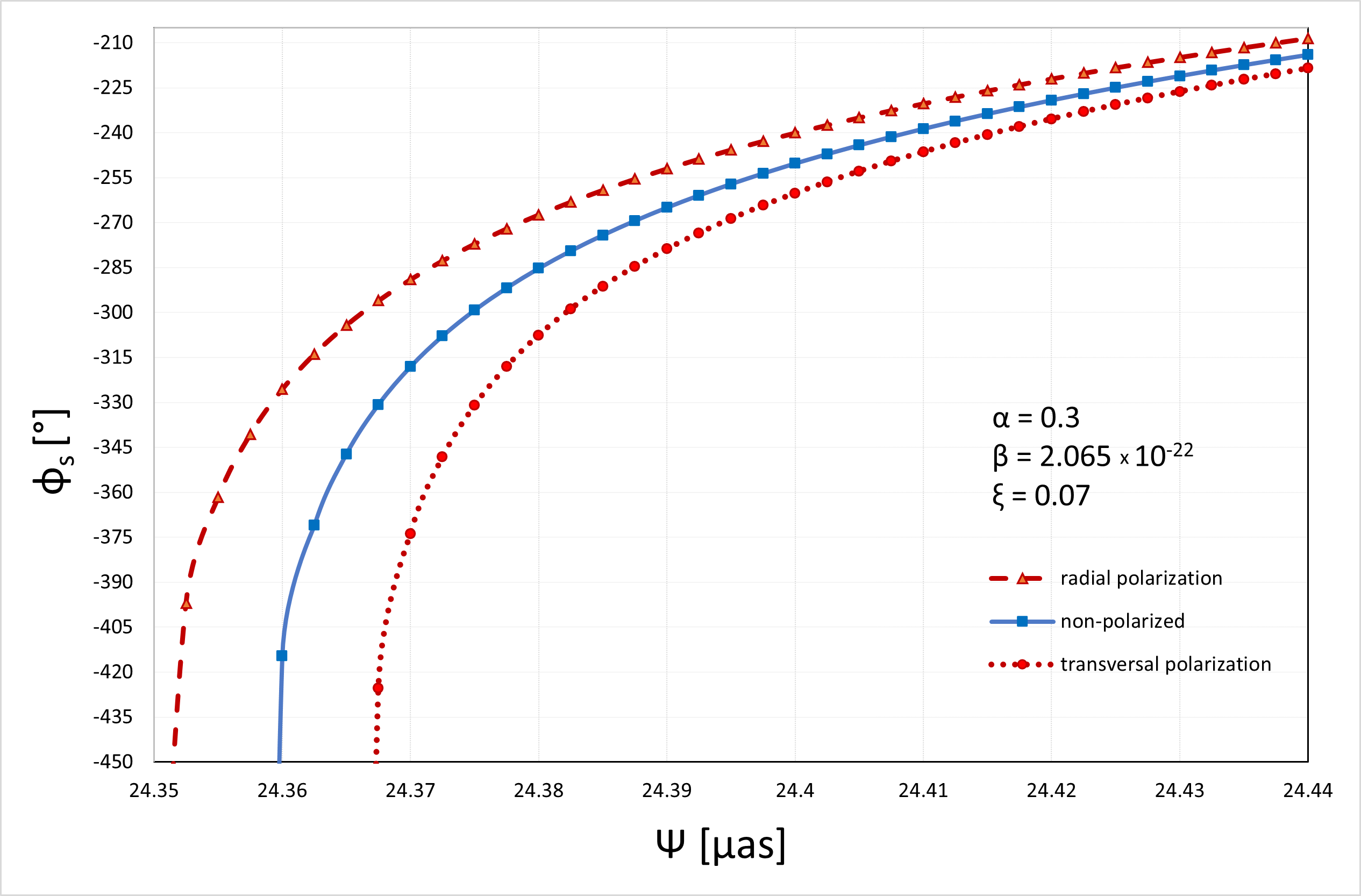}
\caption{\label{figPhiZoom} The angular position $\phi_s$ of a source at $r_s=6M$, as a function of the image angle $\Psi$ for Sagittarius A* with $M=4.154\times 10^{6}M_{\odot}$. The bold line corresponds to the generalized q-metric with values of $\alpha$ and $\beta$ that match the observed angular radius of the shadow (see Figure \ref{figBetas}). The dashed represents photons with radial polarization, while the dotted line represents those with transversal polarization. The plot focuses on the region near the angular shadow ($\Psi \sim 24.35\ \mu as$), where the Schwarzschild case corresponds to larger image angles  ($\Psi\sim 26.06\ \mu as$, as shown  in Fig. \ref{figPhi}). }

\end{figure*}

Figures \ref{figPhi} and \ref{figPhiZoom} illustrate the angular position $\phi_s$ of sources at celestial angles $\Psi$, as predicted by the geodesic motion of photons in the backwards ray-tracing method, specifically for the case of Sagittarius A* with a source at $r_s=6M$. The distance from Earth expressed in terms of the mass of Sagittarius A* ($M = 4.154 \times 10^6 M_{\odot}$), is approximately $r_0 \sim 4.113 \times 10^{10}M$ in geometrized units. The values of $\alpha$ and $\beta$ are chosen such that the critical angle matches the angular shadow radius of Sag A*, $\Psi_c \sim 24.35 \mu as$. With these $\alpha$ and $\beta$ values, we compute $\phi_s$ for photons with radial and transverse polarizations, as well as for unpolarized photons. Figure \ref{figPhi} presents the cases of pure deformation ($\beta=0$) on the right side and pure distortion ($\alpha=0$) on the left side, allowing us to examine the effects of these two parameters separately. The parameter values are selected so that $\Psi_c$ matches the angular radius of the emission ring, complementing Figure \ref{figSagA}. It is clear that the difference between the position $\phi_s$ of the source as predicted by the generalized q-metric and the one by the Schwarzschild solution is significant. On the other hand, observing birefringence effects would require a higher resolution, as it is presented in Figure \ref{figPhiZoom}. Figure \ref{figPhiZoom} shows the angular position of the source $\phi_s$ as a function of the image angle $\Psi$. The vertical asymptotes of each curve correspond to the respective critical angle \eqref{Psic2}, representing the scenario where a light ray reaches the light ring, circles multiple times, and eventually escapes to an observer at the Earth's distance \footnote{We note that this study is implemented in the equatorial plane, while the Earth is not perfectly aligned with the equatorial plane of Sgr A*. In principle, this deviation can affect observed angles such as inclination and position angle, and the observed image of the accretion disk or jet might appear slightly tilted or distorted. This effect is more significant for high-resolution images. However, determining the exact inclination angle between the Earth's line of sight and the equatorial plane of Sgr A* is complicated. Based on both theoretical calculations and observational evidence, assuming alignment introduces minimal error in the calculations for most practical purposes. Given the small inclination angle and the large distance involved, this is a meaningful and reliable approximation for our study.}. We note that in Figure \ref{figPhiZoom} we are focusing on the region closer to the angular shadow $\Psi\sim 24.35\ \mu as$. The Schwarzschild case, corresponding to larger image angles of $\Psi\sim 26.06\ \mu as$  is not shown in these plots to avoid cluttering, as it would overlap with the other curves. This case is included in Figure \ref{figPhi}, where a broader range of image angles is considered for comprehensive comparison.

The polarization effect can be explained as follows: In the right-hand side of Figure \ref{figPhiZoom}, a larger value of $\xi=0.07$ was chosen. For a source positioned at $\phi_s = 315^{\circ}$, a radially polarized image is observed at an angle of $\Psi \sim 24.362 \mu as$, a transversely polarized image at $\Psi \sim 24.379 \mu as$, and a main image (composed of unpolarized photons) at $\Psi \sim 24.37 \mu as$. This illustrates the birefringence effect on the image of the observed source. Additionally, higher-order images of both the main and polarized images appear at angles closer to the angular shadow radius $\Psi_c$, since angles $\phi_s = -675^{\circ}, -1035^{\circ}, ...$ correspond to the same source. A similar effect is observed for negative values of the image angle, $-\pi/2 \leq \Psi \leq 0$, which correspond to positive values of $\phi_s$ and light rays passing through the opposite side of the lens. Furthermore, the birefringence effect would also be visible on the shadow, but observing such effect requires a higher resolution. For instance, with $\xi \sim 0.05$, a resolution of approximately $0.01 \mu as$ is necessary, while for $\xi \sim 0.07$, a resolution of around $0.02 \mu as$ is required.

\section{Summary and conclusion}\label{sum}

In summary, we considered the impact of one-loop vacuum polarization on the photon propagation on the background of a distorted, deformed compact object characterized by quadrupoles. The compact object’s deformation is described by the deformation parameter $\alpha$, while the distortion parameter $\beta$ is related to an additional external gravitational field, like an external mass distribution or a magnetic surrounding. We conducted a detailed analysis of the dependence of the shadow on model parameters. By calculating parameter values that correspond to the observational data of Sgr A*, we explored different scenarios. Additionally, utilizing these observations allows us to refine the valid parameter range of the metric, enhancing its applicability to astrophysical contexts.

Electromagnetic birefringence, which involves the dependence of photon velocity on polarization in the presence of a background electromagnetic field, shows distinct characteristics for different polarization states. For photons with radial and transverse polarization states in radial motion, both the light cone and photon velocity remain unchanged. However, for orbital photons, the velocity can depend on the polarization direction. We identify specific directions and polarizations where the photon velocity exceeds the speed of light, \(c\). Additionally, gravitational effects can increase the velocity for certain directions and polarizations, highlighting the anisotropy of the background field. Furthermore, a nonminimal coupling of gravity and electrodynamics would produce the same effects in the propagation of photons coupled to the Weyl tensor. In spite of the fact that the resolution needed to visualize QED vacuum polarization effects on the images may not be reachable, birefringece effects due to this nonminimal coupling may be measurable for future higher precision observations. In QED, the vacuum is predicted to exhibit birefringence in the presence of a magnetic field, a phenomenon yet to be experimentally confirmed. If this birefringence occurs, the polarization of photons emitted from the accretion disk would alter as they traverse the magnetized vacuum. Further research is required to quantify this effect for photons emerging from the accretion disk plane.

\section*{Acknowledgements}
The authors acknowledge fruitful discussion with Claus
Laemmerzahl and Christian Pfeifer. D.A. acknowledges financial support from the Deutscher Akademischer Austauschdienst (DAAD, German Academic Exchange Servie) fellowship Ref. No. 91832671. Sh.F. acknowledges the University off
Waterloo, the Government of Canada through the Department of Innovation, Science and Economic Development and by the Province of Ontario through the Ministry of Colleges and Universities at Perimeter Institute, also Center of Applied Space Technology and Microgravity (ZARM), University of Bremen, University of Bremen.

\appendix
\section{The Riemann tensor in the tetrad formalism}\label{AppendixA}
Consider the following spherically symmetric spacetime metric:
\begin{equation}\label{metricg}
ds^2=-e^{2T(x,y)}dt^2+e^{2X(x,y)}dx^2+e^{2Y(x,y)}dy^2+e^{2\Phi(x,y)}d\phi^2\, .
\end{equation}
The orthonormal frame, $ds^2=\eta_{ab}\omega^{a}\omega^{b}$
with the Minkowski metric $\eta_{ab}=diag\{-1,1,1,1\}$, is described by the 1-forms
\begin{equation}
\omega^{0}=e^T dt\, ,\quad \omega^{1}=e^X dx\, ,\quad
\omega^{2}=e^Y dy\, ,\quad \omega^{3}=e^\Phi d\phi\, .\label{tetradg}
\end{equation}
The connection forms $\omega^{a}_{\ b}$ are determined as solutions of
the torsion-free equation
\begin{equation}\label{torsiong}
d\omega^{a}+\omega^{a}_{\ c}\wedge\omega^{c}=0\, ,
\end{equation}
with $d$ the exterior derivative of $\omega^{a}$, and 
\begin{equation}\label{conng}
d g_{ab}= \omega_{ab}+\omega_{ba}\, .
\end{equation}
For an orthonormal frame, $g_{ab}=\eta_{ab}$, the derivative $d\eta_{ab}$
vanishes and equation (\ref{conng}) reduces to $\omega_{ab}=-\omega_{ba}$,
which implies $\omega^{a}_{\ a}=0$. With these symmetries, 
the non-vanshing connection forms read as
\begin{eqnarray}
\omega^{0}_{\ 1}&=&e^{-X}T_x\ \omega^{0}\, ,\nonumber\\
\omega^{0}_{\ 2}&=&e^{-Y}T_y\ \omega^{0}\, ,\nonumber\\
\omega^{1}_{\ 2}&=&e^{-Y}X_y\ \omega^{1}-e^{-X}Y_x\ \omega^{2}\, ,\nonumber\\
\omega^{1}_{\ 3}&=&-e^{-X}\Phi_x\ \omega^{3}\, ,\nonumber\\
\omega^{2}_{\ 3}&=&-e^{-Y} \Phi_y\ \omega^{3}\, ,
\label{wab}
\end{eqnarray}
where the subindices denote the partial derivatives of the metric funtions 
(e.g., $T_x\equiv\partial T/\partial x$).

The Riemann curvature 2-form in the orthonormal frame is obtained from the additional
Cartan's equation
\begin{equation}\label{Cartang}
R^{a}_{\ b}=d\omega^{a}_{\ b}+\omega^{a}_{\ c}\wedge\omega^{c}_{\ b}\, .
\end{equation}
In order to identify the components $R^{ab}_{\ \ cd}$ of the Riemann tensor
in the orthonormal frame $\omega^{a}$, we use the definition of the Riemann 2-form
\begin{equation}
R^{ab}=R^{ab}_{\ \ \vert cd\vert}\omega^{c}\wedge\omega^{d}\, .
\end{equation}
The nonvanishing components of the Riemann tensor in the orthonormal frame read as
\begin{eqnarray}
R^{01}_{\ \ 01}&=&-\left\{e^{-2X}\left[T_{xx}+T_x(T_x-X_x)\right]
+e^{-2Y}T_y X_y\right\}\, \nonumber\\ 
R^{01}_{\ \ 02}&=& -e^{-(X+Y)}\left[T_{xy}+T_x(T_y-X_y)-T_y Y_x\right]
\, ,\nonumber\\
R^{02}_{\ \ 01}&=& -e^{-(X+Y)}\left[T_{xy}+T_y(T_x-Y_x)-T_x X_y\right]
\, ,\nonumber\\
R^{02}_{\ \ 02}&=&-\left\{e^{-2Y}\left[T_{yy}+T_y(T_y-Y_y)\right]
+e^{-2X}T_x Y_x\right\}\, ,\nonumber\\
R^{03}_{\ \ 03}&=&-\left[e^{-2X}T_x \Phi_x +e^{-2Y}T_y \Phi_y \right]
\, ,\nonumber\\
R^{12}_{\ \ 12}&=&-\left\{e^{-2Y}\left[X_{yy}+X_y(X_y-Y_y)\right]\right.\nonumber\\
&&\left.\qquad
+e^{-2X}\left[Y_{xx}+Y_x(Y_x-X_x)\right]\right\}\, ,\nonumber\\
R^{13}_{\ \ 13}&=&  -\left\{e^{-2X}\left[\Phi_{xx}+\Phi_x(\Phi_x-X_x)\right] 
+ e^{-2Y}\Phi_y X_y\right\}\, \nonumber\\
%
%
R^{13}_{\ \ 23}&=&-e^{-(X+Y)}\left[\Phi_{xy}+\Phi_x(\Phi_y-X_y)-\Phi_y Y_x\right]
\, ,\nonumber\\
R^{23}_{\ \ 13}&=& -e^{-(X+Y)}\left[\Phi_{xy}+\Phi_y(\Phi_x-Y_x)-\Phi_x X_y\right]
\, ,\nonumber\\
R^{23}_{\ \ 23}&=& -\left\{e^{-2Y}\left[\Phi_{yy}+\Phi_y(\Phi_y-Y_y)\right] 
+ e^{-2X}\Phi_x Y_x\right\}\, ,
\label{Rabcdg}
\end{eqnarray}
with the symmetries $R^{ab}_{\ \ cd}=-R^{ba}_{\ \ cd}
=R^{ba}_{\ \ dc}=-R^{ab}_{\ \ dc}$. 
The components of the Riemann tensor in the basis 1-forms $dx^{\mu}$,
are obtained by the transformation relation
\begin{equation}\label{Rmnstg}
R^{\mu\nu}_{\ \ \sigma\tau}=e^{\mu}_{a}e^{\nu}_{b}e_{\sigma}^{c}e_{\tau}^{d}
R^{ab}_{\ \ cd}\, ,
\end{equation}
where $e^{a}_{\mu}$ are defined from $\omega^{a}=e^{a}_{\mu}dx^{\mu}$.

A simple example would be the Reissner-Nordström anti-de Sitter spacetime,
described by equation (\ref{metricg}) for the coordinates $x=r$ and $y=\theta$, and with the metric functions defined by
$T(r)=-X(r)=\frac12\ln\left(1-\frac{2M}{r}+\frac{Q^2}{r^2}-\frac{\Lambda}{3}\right)$,
$Y(r)=\frac12\ln\left(r^2\right)$, and $\Phi(r,\theta)=\frac12\ln\left(r^2\sin^2\theta\right)$. The non-vanishing components of the Riemann tensor in the orthonormal frame equation (\ref{Rabcdg}),
read
\begin{eqnarray}
R^{01}_{\ \ 01}&=&\left(\frac{Q^2}{r^4}-\frac{M}{r^3}
+\frac{\Lambda}{3}\right)-\frac{4Q^2}{r^4}+\frac{3M}{r^3}
\, \nonumber\\ 
R^{23}_{\ \ 23}&=&\left(\frac{Q^2}{r^4}-\frac{M}{r^3}
+\frac{\Lambda}{3}\right)-\frac{2Q^2}{r^4}+\frac{3M}{r^3}\, ,\nonumber\\ 
R^{02}_{\ \ 02}&=&R^{03}_{\ \ 03}=R^{12}_{\ \ 12}= R^{13}_{\ \ 13}
=\frac{Q^2}{r^4}-\frac{M}{r^3}+\frac{\Lambda}{3}\, ,\qquad
\label{RabcdRNL}
\end{eqnarray}
and the Riemann tensor can be rewritten in terms of equation (\ref{Umunu}), as
\begin{eqnarray}
R^{\mu\nu}_{\ \ \sigma\tau}&=&-\left(\frac{M}{r^3}-\frac{Q^2}{r^4}-\frac{\Lambda}{3}\right)
\left[\delta^{\mu}_{\sigma}\delta^{\nu}_{\tau}-\delta^{\mu}_{\tau}\delta^{\nu}_{\sigma}\right]
\, \nonumber\\
&&+\left(\frac{3M}{r^3}-\frac{4Q^2}{r^4}\right)U^{\mu\nu}_{01}U_{\sigma\tau}^{01}
+\left(\frac{3M}{r^3}-\frac{2Q^2}{r^4}\right)
U^{\mu\nu}_{23}U_{\sigma\tau}^{23}\, .\nonumber\\
\end{eqnarray}

\bibliographystyle{unsrt}
\bibliography{namebib}

\newcommand{\noop}[1]{}
\begin{thebibliography}{10}

\bibitem{PhysRevD.22.343}
I.~T. Drummond and S.~J. Hathrell.
\newblock Qed vacuum polarization in a background gravitational field and its
  effect on the velocity of photons.
\newblock {\em Phys. Rev. D}, 22:343--355, Jul 1980.

\bibitem{1981PThPh..65.1058O}
Y.~{Ohkuwa}.
\newblock {Effect of a Background Gravitational Field on the Velocity of
  Neutrinos}.
\newblock {\em Progress of Theoretical Physics}, 65(3):1058--1067, March 1981.

\bibitem{1994NuPhB.425..634D}
R.~D. {Daniels} and G.~M. {Shore}.
\newblock {``Faster than light'' photons and charged black holes}.
\newblock {\em Nuclear Physics B}, 425(3):634--650, August 1994.

\bibitem{1996PhLB..367...75D}
R.~D. {Daniels} and G.~M. {Shore}.
\newblock {``Faster than light'' photons and rotating black holes}.
\newblock {\em Physics Letters B}, 367:75--83, February 1996.

\bibitem{1998NuPhB.524..639C}
Rong-Gen {Cai}.
\newblock {Propagation of vacuum polarized photons in topological black hole
  spacetimes}.
\newblock {\em Nuclear Physics B}, 524(3):639--657, August 1998.

\bibitem{2014PhRvD..89j4014C}
Songbai {Chen} and Jiliang {Jing}.
\newblock {Rotating charged black hole with Weyl corrections}.
\newblock {\em \prd}, 89(10):104014, May 2014.

\bibitem{JING2016219}
Jiliang Jing, Songbai Chen, and Qiyuan Pan.
\newblock Geometric optics for a coupling model of electromagnetic and
  gravitational fields.
\newblock {\em Annals of Physics}, 367:219--226, 2016.

\bibitem{10.1093/mnras/stw2798}
R.~P. Mignani, V.~Testa, D.~González~Caniulef, R.~Taverna, R.~Turolla,
  S.~Zane, and K.~Wu.
\newblock {Evidence for vacuum birefringence from the first optical-polarimetry
  measurement of the isolated neutron star RX J1856.5-3754}.
\newblock {\em Monthly Notices of the Royal Astronomical Society},
  465(1):492--500, 11 2016.

\bibitem{2015JCAP...10..002C}
Songbai {Chen} and Jiliang {Jing}.
\newblock {Strong gravitational lensing for the photons coupled to Weyl tensor
  in a Schwarzschild black hole spacetime}.
\newblock {\em \jcap}, 2015(10):002--002, October 2015.

\bibitem{2016EPJC...76..357L}
Xu~{Lu}, Feng-Wei {Yang}, and Yi~{Xie}.
\newblock {Strong gravitational field time delay for photons coupled to Weyl
  tensor in a Schwarzschild black hole}.
\newblock {\em European Physical Journal C}, 76(7):357, July 2016.

\bibitem{2018EPJC...78..191C}
Wei-Guang {Cao} and Yi~{Xie}.
\newblock {Weak deflection gravitational lensing for photons coupled to Weyl
  tensor in a Schwarzschild black hole}.
\newblock {\em European Physical Journal C}, 78(3):191, March 2018.

\bibitem{2020PhRvD.101l4038B}
Santiago E.~P. {Bergliaffa}, Edson Elias de~Souza {Filho}, and Rodrigo {Maier}.
\newblock {Strong lensing and nonminimally coupled electromagnetism}.
\newblock {\em \prd}, 101(12):124038, June 2020.

\bibitem{2018AnPhy.399..193O}
Ali {{\"O}vg{\"u}n}, Kimet {Jusufi}, and {\.I}zzet {Sakall{\i}}.
\newblock {Gravitational lensing under the effect of Weyl and bumblebee
  gravities: Applications of Gauss-Bonnet theorem}.
\newblock {\em Annals of Physics}, 399:193--203, December 2018.

\bibitem{2020ChPhC..44i5105A}
G.~{Abbas}, Asif {Mahmood}, and M.~{Zubair}.
\newblock {Strong gravitational lensing for photon coupled to Weyl tensor in
  Kiselev black hole}.
\newblock {\em Chinese Physics C}, 44(9):095105, September 2020.

\bibitem{2021EPJC...81..991Z}
Zelin {Zhang}, Songbai {Chen}, Xin {Qin}, and Jiliang {Jing}.
\newblock {Polarized image of a Schwarzschild black hole with a thin accretion
  disk as photon couples to Weyl tensor}.
\newblock {\em European Physical Journal C}, 81(11):991, November 2021.

\bibitem{2023Univ....9..130A}
Ghulam {Abbas}, Ali {\"Ovg\"un}, Asif {Mahmood}, and Muhammad {Zubair}.
\newblock {Strong Deflection Gravitational Lensing for the Photons Coupled to
  the Weyl Tensor in a Conformal Gravity Black Hole}.
\newblock {\em Universe}, 9(3):130, March 2023.

\bibitem{chen2023kerr}
Songbai Chen and Jiliang Jing.
\newblock Kerr black hole shadow casted by the extraordinary light rays with
  weyl corrections.
\newblock {\em Science China Physics, Mechanics \& Astronomy}, 67:250411, 2024.

\bibitem{2000PhRvD..62h4003V}
K.~S. {Virbhadra} and George F.~R. {Ellis}.
\newblock {Schwarzschild black hole lensing}.
\newblock {\em \prd}, 62(8):084003, October 2000.

\bibitem{2002PhRvD..66j3001B}
V.~{Bozza}.
\newblock {Gravitational lensing in the strong field limit}.
\newblock {\em \prd}, 66(10):103001, November 2002.

\bibitem{PhysRevD.59.124001}
Simonetta Frittelli and Ezra~T. Newman.
\newblock Exact universal gravitational lensing equation.
\newblock {\em Phys. Rev. D}, 59:124001, Apr 1999.

\bibitem{PhysRevD.61.064021}
Simonetta Frittelli, Thomas~P. Kling, and Ezra~T. Newman.
\newblock Spacetime perspective of schwarzschild lensing.
\newblock {\em Phys. Rev. D}, 61:064021, Feb 2000.

\bibitem{Note1}
However, this is not in favor of the primeval fields scenario and therefore
  inflation which is a prime candidate for the production of primeval magnetic
  fields. This scenario requires $r > 10^{-10}$ for the scales of astrophysical
  interest, $\lambda $ around Mpc. However, considering the mentioned approach
  the primeval fields produced are typically small $r \sim 10^{-68}$.

\bibitem{doi:10.1002/andp.19173591804}
Hermann Weyl.
\newblock Zur gravitationstheorie, 1917.

\bibitem{universe8030195}
Shokoufe Faraji.
\newblock Circular geodesics in a new generalization of q-metric.
\newblock {\em Universe}, 8(3), 2022.

\bibitem{1976JMP....17...54E}
F.~J. {Ernst}.
\newblock {Black holes in a magnetic universe}.
\newblock {\em Journal of Mathematical Physics}, 17(1):54--56, 1976.

\bibitem{Note2}
An alternative way of deriving the dispersion relations is to start from the
  wave equation in the radiation gauge, where there are components only in the
  transverse plane, with zero time-like and longitudinal components. The wave
  equation reduces to the 2-dimensional equation \cite {Melrose2008}.

\bibitem{1966MNRAS.131..463S}
J.~L. {Synge}.
\newblock {The escape of photons from gravitationally intense stars}.
\newblock {\em \mnras}, 131:463, January 1966.

\bibitem{Nickalls_1993}
R.W.D. Nickalls.
\newblock A new approach to solving the cubic: Cardan’s solution revealed.
\newblock {\em The Mathematical Gazette}, 77(480):354–359, 1993.

\bibitem{2019A&A...625L..10G}
{GRAVITY Collaboration}.
\newblock {A geometric distance measurement to the Galactic center black hole
  with 0.3\% uncertainty}.
\newblock {\em \aap}, 625:L10, May 2019.

\bibitem{2022ApJ...930L..12E}
{Event Horizon Telescope Collaboration}.
\newblock {First Sagittarius A* Event Horizon Telescope Results. I. The Shadow
  of the Supermassive Black Hole in the Center of the Milky Way}.
\newblock {\em \apjl}, 930(2):L12, May 2022.

\bibitem{2002Natur.419..694S}
{Sch{\"o}del} et~al.
\newblock {A star in a 15.2-year orbit around the supermassive black hole at
  the centre of the Milky Way}.
\newblock {\em \nat}, 419(6908):694--696, October 2002.

\bibitem{2021arXiv210213000G}
Reinhard {Genzel}.
\newblock {A Forty Year Journey}.
\newblock {\em arXiv e-prints}, page arXiv:2102.13000, February 2021.

\bibitem{1998ApJ...509..678G}
A.~M. {Ghez}, B.~L. {Klein}, M.~{Morris}, and E.~E. {Becklin}.
\newblock {High Proper-Motion Stars in the Vicinity of Sagittarius A*: Evidence
  for a Supermassive Black Hole at the Center of Our Galaxy}.
\newblock {\em \apj}, 509(2):678--686, December 1998.

\bibitem{2019ApJ...875L...1E}
{Event Horizon Telescope Collaboration}.
\newblock {First M87 Event Horizon Telescope Results. I. The Shadow of the
  Supermassive Black Hole}.
\newblock {\em \apjl}, 875(1):L1, April 2019.

\bibitem{Note3}
We note that this study is implemented in the equatorial plane, while the Earth
  is not perfectly aligned with the equatorial plane of Sgr A*. In principle,
  this deviation can affect observed angles such as inclination and position
  angle, and the observed image of the accretion disk or jet might appear
  slightly tilted or distorted. This effect is more significant for
  high-resolution images. However, determining the exact inclination angle
  between the Earth's line of sight and the equatorial plane of Sgr A* is
  complicated. Based on both theoretical calculations and observational
  evidence, assuming alignment introduces minimal error in the calculations for
  most practical purposes. Given the small inclination angle and the large
  distance involved, this is a meaningful and reliable approximation for our
  study.

\bibitem{Melrose2008}
D.~B. Melrose.
\newblock {\em Quantum plasmadynamics: unmagnetized plasmas}.
\newblock Springer, 2008.

\end{thebibliography}
\end{document}